\documentclass[11pt]{article}
\usepackage[letterpaper,margin=0.87in]{geometry}
\usepackage{amsmath,amssymb,amsthm,graphicx,booktabs,microtype,placeins,array}
\usepackage[T1]{fontenc}
\usepackage{lmodern}
\usepackage[colorlinks=true,linkcolor=blue,citecolor=blue,urlcolor=blue]{hyperref}
\hypersetup{pdftitle={Network imitation sustains misinformation despite a corrective factual field},pdfauthor={Ruiwu Niu}}
\usepackage{caption}
\newtheorem{theorem}{Theorem}
\newtheorem{proposition}[theorem]{Proposition}
\newcommand{\R}{\mathbb R}
\newcommand{\uvec}{\mathbf u}
\newcommand{\avec}{\mathbf a}
\newcommand{\fvec}{\mathbf f}
\newcommand{\bvec}{\mathbf b}
\newcommand{\K}{\mathcal K}
\newcommand{\E}{\mathcal E}
\DeclareMathOperator{\diag}{diag}
\title{Network imitation sustains misinformation\\despite a corrective factual field}
\author{Ruiwu Niu\thanks{\href{mailto:rniu@hksyu.edu}{rniu@hksyu.edu}}\\[0.4em]
\small Department of Data Science and Digital Innovation\\
\small Hong Kong Shue Yan University, Hong Kong SAR, China}
\date{20 September 2026}
\begin{document}
\maketitle
\begin{abstract}
Social influence can preserve errors that individuals would correct on their own. We study this possibility by representing opinions as directions in a semantic space. The formulation connects established models of personal preference, social agreement, and external influence. A common factual signal pulls every opinion toward a reference direction, individuals remain attached to their initial preferences, and connected individuals imitate one another. When the factual signal exceeds the preference strength, each isolated individual corrects, apart from an exactly opposite initial opinion. We prove that every finite undirected network then has a unique state of minimum total energy in which all opinions favor the factual direction. This state is the best overall compromise between factual alignment, personal preferences, and agreement with neighbors. Yet on a ring with planar opinions, sufficiently strong imitation can trap the same individuals in a different stable state that retains misinformation. We prove convergence to that state from the specified initial opinions. We also derive a network spectral condition that excludes competing stable states, and show how an additional semantic direction can destabilize the constructed planar pattern. Numerical maps reveal persistent error, history dependence, and coexisting outcomes, while comparisons across networks show differences that a single spectral scale does not remove. These results distinguish individual correctability from collective correction and identify how attractive social influence can obstruct the latter.
\end{abstract}

\section{Introduction}
Agreement and factual accuracy are different properties of a collective opinion. A group can agree on an incorrect account, and increasing agreement need not improve its relation to the evidence. The persistence of misinformation after correction makes this distinction a practical as well as a mathematical concern \cite{Ecker2022}. Models of information diffusion describe how messages spread, while opinion models describe how individuals change what they believe. The distinction matters when people receive both a persistent factual signal and repeated social reinforcement. We examine whether individuals who would correct their opinions in isolation can retain errors after they begin imitating their neighbors.

We represent each opinion by a unit vector in a semantic space. Its angle to a common factual direction measures its departure from that reference. Each individual is influenced by the factual signal, a fixed preferred direction, and neighboring opinions. The preferred direction is also the initial opinion. This choice allows a controlled comparison in which the individuals, their preferences, and their initial states remain identical, and only the strength of social interaction changes. The network is held fixed during each run; its structure is varied between experiments.

Our first contribution is the integration of these influences within a common spherical framework. Exact parameter and graph reductions connect it to social compass dynamics, spherical consensus, and forced planar opinion models, while its energy connects it to vector spin systems with fixed local fields. The integration supplies a common factual observable and a consistent comparison in which preferences and initial opinions remain fixed as interactions change.

The main theoretical result is a separation between the best collective compromise and the outcome reached by the dynamics. When the factual signal is stronger than the individual preference, every finite network has a unique global energy minimum in which all opinions have a positive factual projection. The energy combines disagreement with the factual signal, disagreement with the fixed preference, and disagreement between neighbors. Its minimum therefore describes the state that best satisfies all three influences together. Nevertheless, a ring network can converge to a higher local minimum containing incorrect opinions. The same individuals correct when disconnected. Persistent misinformation can thus arise from the path taken under mutual influence even when the global minimum favors the factual direction at every node.

The mathematical contribution is to prove this selection from specified initial opinions. Establishing an incorrect equilibrium alone would leave open whether the stated model ever reaches it. We construct an initial pattern on a finite ring, derive a sufficient interaction strength, and show that energy decreases within a region from which the trajectory cannot escape. A separate argument establishes the unique factual global minimum on any finite graph. Together these results demonstrate that individual correctability and a factual global minimum do not guarantee collective correction.

The analysis also explains when competing stable states can be excluded and when a particular incorrect state loses stability. A bound involving the signless graph Laplacian gives a region with only one stable equilibrium. For the constructed planar pattern, the full stability matrix separates motion within the plane from motion out of it. The latter can destabilize a state that is stable in the plane. Numerical experiments then examine randomly sampled preferences, history dependence, and the effects of graph structure, semantic dimension, and population size. Their role is to show how the mechanism appears beyond the explicit construction and to identify which variations require further theory.

\section{Related opinion and spin models}\label{sec:related}
Persistent misinformation has several possible sources, including cognitive factors and the social environment in which corrections are received \cite{Ecker2022}. Formal truth seeking models have long combined an external truth value with exchange among individuals \cite{Hegselmann2006}. Friedkin and Johnsen established a different ingredient, continuing attachment to an initial opinion during social influence \cite{Friedkin1990}. These contributions motivate the two fixed influences in the present model. The nonlinear constraint to unit opinion vectors makes the dynamics different from affine opinion averaging.

Directional opinion models provide the closest mathematical setting. Caponigro, Lai, and Piccoli developed opinion formation on a sphere with external media influence \cite{Caponigro2015}. The social compass model combines attraction toward initial preferences with imitation, first for planar opinions and then for multidimensional opinions on networks \cite{Ojer2023,Ojer2025}. Sampson and Restrepo obtained reduced dynamics and community extensions \cite{SampsonRestrepo2026}; their conference abstract and public implementation also include external broadcasts \cite{SampsonRestrepoPoster2025,SampsonCode2026}. The implementation begins its first coupling sweep at the preferred directions. It therefore supplies an explicit planar precedent for the three forces and their connection to the initial state. Recent spherical models also produce multiple opinion clusters by modifying the interaction law \cite{Zhang2025}. The present analysis retains ordinary attractive alignment and studies the competition with fixed personal and factual influences.

Other recent models distinguish an internal belief from the opinion expressed to others. Riazi and Livan examine public and private beliefs under disinformation, and compare strategies for correcting false information \cite{Riazi2024,Riazi2026}. Luo and colleagues study how a conflict between social pressure and perceived truth can produce pluralistic ignorance through Pareto improvement dynamics \cite{Luo2026}. Our single opinion vector does not introduce strategic deception or separate public and private states. Instead, it isolates a mechanism through which attractive interaction can sustain error even when the factual signal acts continuously on every individual.

The energy formulation connects this problem to established spin models. A ferromagnetic vector spin system rewards alignment between neighboring unit vectors and with fixed local fields. Uniform and random local fields have been studied in the XY model \cite{Garanin2013}, including on sparse random graphs \cite{Lupo2019}. Finite XY chains with pinned ends can support metastable helices \cite{PopovPini2018} and a response to weak applied fields \cite{Anisimov2019}. Related phase waves and their entrainment occur on driven Kuramoto rings \cite{RoyLahiri2012}. The difference between stability on a circle and on a sphere is also established in consensus theory \cite{Markdahl2018}. These precedents identify the relevant geometry and energy mechanisms. Here we prove the combination of a unique factual global minimum and convergence to an incorrect local minimum from the same specified initial opinions that correct in isolation.

The role of network structure is related to earlier work by Niu, Shu, and Zhao on selection of opposite collective opinions in networks with identical Laplacian eigenvalues \cite{NiuShuZhao2026}. That study uses contextual message processing and a different nonlinear dynamics. The present work instead asks how factual orientation, local attraction, and the selected configuration enter stability. It supplies a sufficient spectral bound while showing why that bound cannot determine all outcomes outside its guaranteed region.

\section{Opinions, factual influence, and imitation}\label{sec:model}
\subsection{Semantic directions and a separate influence network}
Consider $N$ individuals, where $N$ is a positive integer. Individuals are indexed by $i\in\{1,\ldots,N\}$, and time is $t\in[0,\infty)$. The semantic dimension is an integer $n\geq2$. Individual $i$ has a current opinion $\uvec_i(t)\in S^{n-1}$, where
\begin{equation}
 S^{n-1}=\{\mathbf v\in\R^n\mid\|\mathbf v\|=1\}
\end{equation}
is the unit sphere in $\R^n$ and $\|\cdot\|$ is the Euclidean norm. Opinion length is fixed, so the model describes direction rather than confidence or message volume. The vectors $\fvec\in S^{n-1}$ and $\avec_i\in S^{n-1}$ denote the factual reference and the individual's fixed preferred direction. Both are constant in time. We also call $\avec_i$ an anchor for the current opinion. This is attraction to a persistent personal preference, often parameterized by conviction in the social compass model \cite{Ojer2023,Ojer2025}. Persistent attachment to initial opinions also has a precedent in affine social influence models \cite{Friedkin1990}.

The factual projection is $q_i(t)=\uvec_i(t)\cdot\fvec\in[-1,1]$, where the dot denotes the Euclidean inner product. Equivalently, $q_i=\cos\vartheta_i$ for the angle $\vartheta_i\in[0,\pi]$ between the opinion and the factual direction. We classify $q_i>0$ as information and $q_i<0$ as misinformation within this model. The boundary $q_i=0$ is orthogonal to the factual direction. A positive projection means that an opinion is closer to $\fvec$ than to $-\fvec$; near agreement with the factual reference requires $q_i$ close to one. Correction below means entry into the positive hemisphere, not necessarily exact alignment.

These definitions assign a geometrical meaning to factual orientation. The $n-1$ directions perpendicular to $\fvec$ describe different ways of departing from that reference. They are not directions of transmission between individuals. The network determines who influences whom, while the semantic vectors describe the opinions being influenced. Figure~\ref{fig:model}(a,b) separates these two spaces. This theory assumes that the factual reference and semantic directions have been defined for the question being studied. Constructing them from statements or language model representations is a subsequent measurement problem.

\subsection{Three influences rotate the current opinion}
The network is represented by an adjacency matrix $A\in\R^{N\times N}$. We assume $A=A^\top$, $A_{ij}\geq0$, and $A_{ii}=0$. The entry $A_{ij}$ is a dimensionless weight measuring mutual influence between individuals $i$ and $j$; zero means no edge. These are raw weights, without division by node degree. A node with more incident weight can therefore receive a larger total social influence.

The dynamics are
\begin{align}
 \dot{\uvec}_i&=\Gamma P_i\mathbf H_i,
 &P_i&=I_n-\uvec_i\uvec_i^\top, \label{eq:model}\\
 \mathbf H_i&=h\fvec+\kappa\avec_i+J\sum_{j=1}^N A_{ij}\uvec_j.
 \label{eq:field}
\end{align}
Here $I_n$ is the $n\times n$ identity matrix, the superscript $\top$ denotes transpose, and a dot over a vector denotes its time derivative. The factual strength $h\in(0,\infty)$ is uniform across individuals. The preference strength $\kappa\in[0,\infty)$ and interaction strength $J\in[0,\infty)$ are also common to all individuals. They have the same field units. The response coefficient $\Gamma\in(0,\infty)$ converts a field into a rotation rate, so $\Gamma h$ has units of inverse time. Changing $\Gamma$ rescales the speed of every trajectory and leaves its path, equilibria, and stability signs unchanged.

\begin{figure}[!tbp]
\centering
\includegraphics[width=\textwidth]{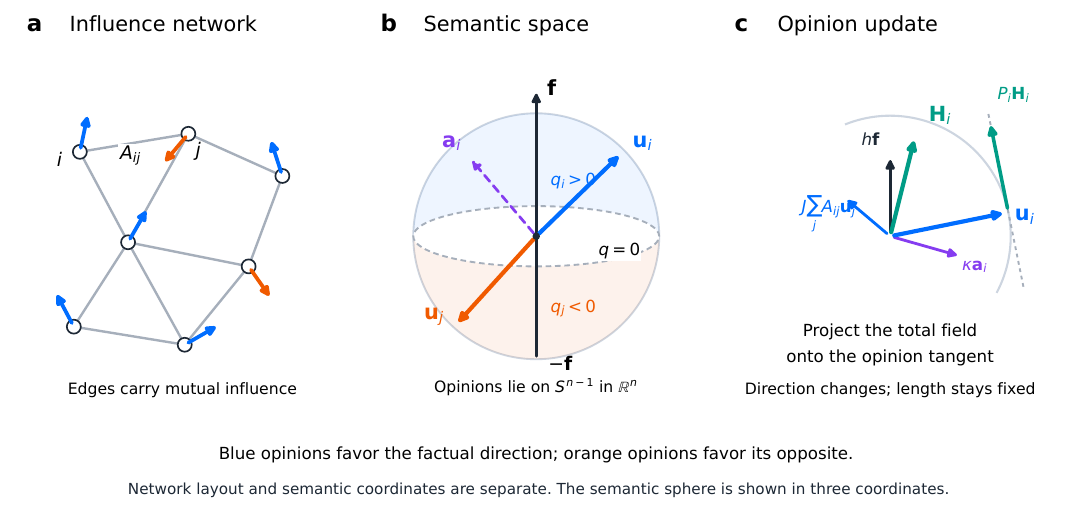}
\caption{\textbf{The network carries influence between opinions in a semantic space.}
(a) Nodes represent individuals and edges specify the weights $A_{ij}$. The arrows attached to nodes are opinions expressed in common semantic coordinates; the placement of nodes has no semantic meaning. (b) All opinion and preference vectors can be placed at the origin of the same $n$ dimensional space. The illustration shows three coordinates. The hyperplane $q=0$ separates positive and negative factual projections, and its other coordinates describe departures from the reference. (c) The factual field, fixed preference, and neighboring opinions contribute to $\mathbf H_i$. Its tangent component $P_i\mathbf H_i$ rotates $\uvec_i$ without changing its length. This is a schematic of the model, not a simulated trajectory.}
\label{fig:model}
\end{figure}

The total field $\mathbf H_i\in\R^n$ is the vector sum of the three influences. The matrix $P_i$ projects that field onto the tangent space perpendicular to $\uvec_i$. Thus only the component that rotates the opinion contributes to its motion, and $\frac{d}{dt}\|\uvec_i\|^2=0$. The factual term rotates the opinion toward $\fvec$, the preference term toward $\avec_i$, and the social term toward neighboring directions. Their instantaneous torques can reinforce or oppose one another. There is no restriction that social influence lie in the hyperplane perpendicular to $\fvec$. Figure~\ref{fig:model}(c) illustrates the projection that turns the total influence into a change of direction.

\subsection{Initial opinions and observable outcomes}
Our main comparison uses
\begin{equation}
 \uvec_i(0)=\avec_i. \label{eq:preparation}
\end{equation}
A preferred direction therefore has two roles. It specifies the initial opinion and continues to attract that opinion later. It remains fixed when the current opinion is perturbed. For random initial conditions, we draw preferred directions independently from the probability distribution
\begin{equation}
 \mu_p=(1-p)\sigma_+ +p\sigma_-,\qquad p\in[0,1]. \label{eq:law}
\end{equation}
The probability measures $\sigma_+$ and $\sigma_-$ are uniform on the open positive and negative hemispheres of $S^{n-1}$, respectively, with total mass one on each hemisphere. Thus $p$ is the probability that a given initial opinion lies in the misinformation hemisphere, rather than an imposed fraction in a finite network. The choice $p=1/2$ is uniform on the whole sphere, while $p=0.1$ favors initially factual opinions. We reserve a separate experiment for varying current initial opinions while holding the preferred directions fixed.

Two observables distinguish the prevalence of error from mean alignment,
\begin{equation}
 M(t)=\frac1N\sum_{i=1}^N\mathbf1_{\{q_i(t)<0\}}\in[0,1],\qquad
 \overline q(t)=\frac1N\sum_{i=1}^N q_i(t)\in[-1,1]. \label{eq:M}
\end{equation}
The indicator $\mathbf1_{\{q_i<0\}}$ equals one for a negative projection and zero otherwise. Hence $M$ is the fraction of individuals holding misinformation, and $\overline q$ is the average factual projection. An opinion on the boundary is counted in neither open hemisphere. A subscript $\infty$ denotes the limiting value at a converged state. Large positive $\overline q$ can conceal a minority of negative opinions. By comparison, the usual coherence $\|N^{-1}\sum_i\uvec_i\|\in[0,1]$ measures agreement but gives the same value for consensus at $\fvec$ and consensus at $-\fvec$.

\subsection{Energy and the isolated individual}
The symmetry of influence gives the model an energy function. Define the fixed local field $\bvec_i=h\fvec+\kappa\avec_i\in\R^n$ and the joint opinion configuration $\uvec=(\uvec_1,\ldots,\uvec_N)\in(S^{n-1})^N$. Its energy is
\begin{equation}
 \E(\uvec)=-\sum_{i=1}^N\bvec_i\cdot\uvec_i
      -J\sum_{i<j}A_{ij}\uvec_i\cdot\uvec_j. \label{eq:energy}
\end{equation}
The minus signs make alignment with each influence lower the energy. Lower energy means a better combined compromise with the factual field, the preferences, and the neighbors. Energy is a potential for the opinion dynamics, not a separate empirical measure of truth. A direct calculation gives
\begin{equation}
 \dot\E=-\Gamma\sum_{i=1}^N\|P_i\mathbf H_i\|^2\leq0.
 \label{eq:dissipation}
\end{equation}
The dynamics are therefore gradient descent constrained to the product of unit spheres. A ground state means a global energy minimum. A strict local minimum can also attract nearby states, even if its energy exceeds that of the ground state. We use metastable state in this latter energy sense. In the deterministic model studied here, a stable local minimum can retain the system indefinitely; no finite escape time is implied.

When $J=0$, each individual rotates independently in the fixed field $\bvec_i$. If $\bvec_i\ne0$, its attracting direction is
\begin{equation}
 \uvec_i^{\rm iso}=\frac{h\fvec+\kappa\avec_i}
 {\sqrt{h^2+\kappa^2+2h\kappa\,\avec_i\cdot\fvec}}. \label{eq:isolated}
\end{equation}
The superscript denotes this isolated equilibrium. If $h>\kappa$, its factual projection is positive for every preferred direction. Starting from Eq.~\eqref{eq:preparation}, the individual reaches it unless $\avec_i=-\fvec$. At this exactly opposite direction the torque is zero, although the equilibrium is unstable. This exception has probability zero under the continuous law in Eq.~\eqref{eq:law}. At $h=\kappa$ with $\avec_i=-\fvec$, the two fixed fields cancel. The central correction results use $h>\kappa$, while the numerical maps also examine stronger preferences.

\subsection{Exact relation to existing directional models}\label{sec:modelmap}
For planar opinions, choose $\fvec=(1,0)$ and write $\uvec_i=(\cos\theta_i,\sin\theta_i)$ and $\avec_i=(\cos\psi_i,\sin\psi_i)$. The angles $\theta_i,\psi_i\in\R/(2\pi\mathbb Z)$ are measured modulo a full turn; $\theta_i$ changes in time and $\psi_i$ is fixed. Equation~\eqref{eq:model} reduces to
\begin{equation}
 \Gamma^{-1}\dot\theta_i=-h\sin\theta_i
 -\kappa\sin(\theta_i-\psi_i)
 +J\sum_j A_{ij}\sin(\theta_j-\theta_i). \label{eq:angles}
\end{equation}
Setting $h=0$ as a limiting model recovers social compass dynamics with the same conviction $\kappa$ for every individual. This relation extends to vector opinions, since multidimensional social compass dynamics on networks are already established \cite{Ojer2023,Ojer2025}. Setting both $h=0$ and $\kappa=0$ recovers the standard attractive spherical consensus dynamics \cite{Markdahl2018}. Retaining only $\kappa=0$ gives spherical alignment in a uniform external field.

There is also a direct forced planar precedent. The public implementation accompanying the work of Sampson and Restrepo includes fixed preferences, external broadcasts, and coupling within and between communities \cite{SampsonRestrepo2026,SampsonRestrepoPoster2025,SampsonCode2026}. In its notation, a broadcast of strength $F\geq0$ and direction $\varphi\in\R/(2\pi\mathbb Z)$ contributes $F\sin(\varphi-\theta_i)$, and $\rho_i\geq0$ is conviction. Choosing $F=h$, $\varphi=0$, and $\rho_i=\kappa$ gives Eq.~\eqref{eq:angles}, after the time change $\widetilde t=\Gamma t$. For a single complete community, its within community coupling $K_{\rm in}\geq0$ is divided by $N$; it matches our unit edge model when $K_{\rm in}=NJ$. Equal community sizes also give symmetric weights. The arbitrary symmetric adjacency used here includes sparse cycles beyond these community averages, and the vector formulation allows motion outside the plane. These relations let us study individual correction and collective persistence with the same variables and observables.

There is a further equivalence at the level of the vector field. Whenever $\bvec_i\ne0$, define an effective conviction $\widehat\rho_i=\|\bvec_i\|>0$ and effective preferred direction $\widehat{\mathbf n}_i=\bvec_i/\|\bvec_i\|\in S^{n-1}$. Then Eq.~\eqref{eq:model} has the multidimensional social compass form with a possibly heterogeneous conviction \cite{Ojer2025}. If $\bvec_i=0$, its effective conviction is zero and the direction of that term is immaterial. This rewriting does not preserve the usual identification of preferred and initial directions, since our initial opinion remains $\avec_i$, generally different from $\widehat{\mathbf n}_i$. Retaining $\fvec$ and $\avec_i$ separately is therefore necessary for the correction observable and the comparison made here. Table~\ref{tab:models} collects these relationships.

\begin{table}[!tbp]
\centering
\small
\caption{Model integration through parameter reductions, field redefinition, and a shared energy. All dynamical mappings use the rescaled time $\widetilde t=\Gamma t$.}
\label{tab:models}
\renewcommand{\arraystretch}{1.18}
\begin{tabular}{@{}>{\raggedright\arraybackslash}p{0.25\textwidth}>{\raggedright\arraybackslash}p{0.70\textwidth}@{}}
\toprule
Related model & Exact relationship and retained information \\
\midrule
Social compass dynamics \cite{Ojer2023,Ojer2025} & At $h=0$, Eq.~\eqref{eq:model} gives the common conviction case, with conviction $\kappa$ and initial directions $\avec_i$. This holds for planar and multidimensional opinions. \\
Forced planar social compass \cite{SampsonRestrepoPoster2025,SampsonCode2026} & At $n=2$, a constant broadcast along $\fvec$, common conviction, and the corresponding symmetric community weights reproduce the three terms in Eq.~\eqref{eq:angles}. General symmetric $A$ also permits sparse graphs. \\
Spherical consensus \cite{Markdahl2018} & At $h=\kappa=0$, only projected attractive neighbor alignment remains. The factual observable can still be defined relative to an externally specified reference. \\
Effective preferred field \cite{Ojer2025} & Replacing the two fixed influences by $\widehat\rho_i\widehat{\mathbf n}_i=\bvec_i$ preserves the vector field. It generally changes the preferred direction relative to the prescribed initial opinion, so the initial condition and factual reference must be retained separately. \\
Vector spin energy \cite{Garanin2013,Lupo2019} & Equation~\eqref{eq:energy} is the ferromagnetic vector spin energy with local fields $\bvec_i=h\fvec+\kappa\avec_i$. Common $h$ and $\kappa$ specify a constrained family of local fields. This is an exact energy correspondence; the present dynamics is its constrained gradient flow. \\
\bottomrule
\end{tabular}
\end{table}
\FloatBarrier

\section{A factual global minimum can coexist with persistent error}\label{sec:ground}
Individual correction might fail collectively because the best compromise itself contains misinformation, or because the dynamics settle in another state. The following result separates these possibilities. Under a factual signal stronger than the preferences, the global minimum cannot be the source of persistent error.

\begin{theorem}[Unique factual ground state]\label{thm:ground}
Suppose $\bvec_i\cdot\fvec>0$ at every node, $A=A^\top$ has nonnegative entries, and $J\geq0$. The energy in Eq.~\eqref{eq:energy} has a unique global minimizer, and every opinion there has positive factual projection. It is the only equilibrium with $q_i>0$ for all $i$ and is locally exponentially stable. The closed region with $q_i\geq0$ for all $i$ is forward invariant. Every trajectory starting in this region enters its interior and converges to that minimizer.
\end{theorem}

The condition holds for all preferred directions if $h>\kappa$, since $\bvec_i\cdot\fvec\geq h-\kappa>0$. The graph may be disconnected, and the interaction strength may be arbitrarily large. Local exponential stability means that sufficiently small perturbations decay at an exponential rate. Heterogeneous preferences can still produce different final directions, so factual correction does not require consensus. Forward invariance means that a group whose opinions all have nonnegative projections cannot create a negative projection through these dynamics. The theorem consequently proves correction for the random initial law with $p=0$.

The proof has two steps with distinct meanings. Reflecting a negative factual component into the positive hemisphere lowers its local field cost and cannot increase the total cost of attractive interactions when all negative components are reflected together. This places every global minimizer in the factual hemisphere. Within that hemisphere, coordinates perpendicular to $\fvec$ turn the energy into a strictly convex function, which permits only one minimum. Appendix~\ref{app:ground} provides the calculation and the convergence argument.

\begin{figure}[!tbp]
\centering
\includegraphics[width=\textwidth]{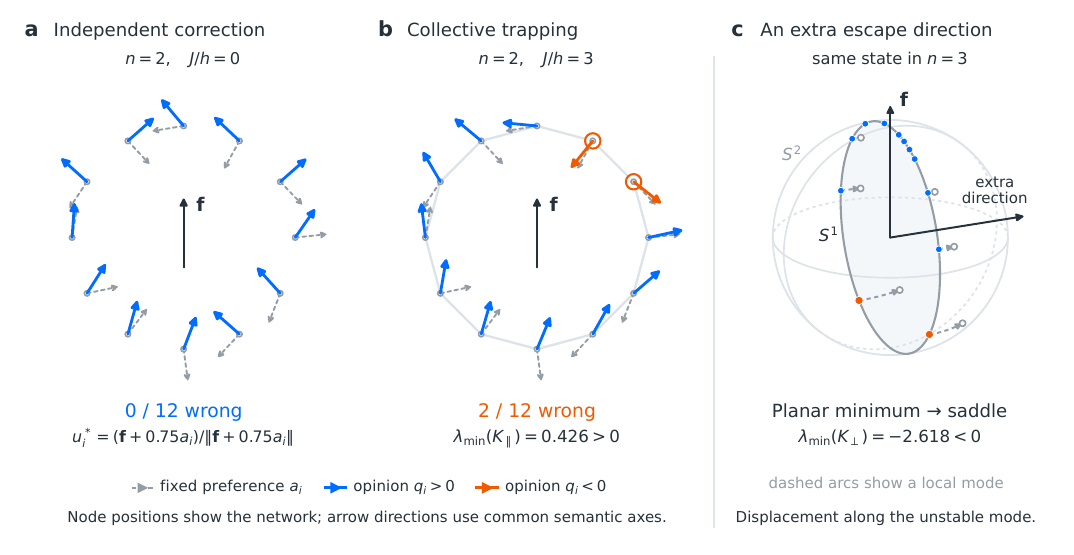}
\caption{\textbf{Individual correction and collective persistence from the same initial opinions.}
Panels (a,b) use the same randomly sampled preferences on $N=12$ nodes, with $h=1$, $\kappa=0.75$, and $\uvec_i(0)=\avec_i$. Disconnected individuals reach $M_\infty=0$. On a ring with unit edge weights and $J=3$, two opinions remain negative. Dashed arrows show fixed preferences and solid arrows show final opinions; all use common semantic coordinates with $\fvec$ upward. Panel (c) places the same planar equilibrium and fixed fields in three semantic dimensions. The dashed paths follow a scaled unstable perturbation out of the plane. They depict a possible local displacement, not an integrated trajectory. Here $\K_\parallel,\K_\perp\in\R^{N\times N}$ are the energy curvature matrices for motion within and out of the opinion plane, and $\lambda_{\min}$ denotes their smallest eigenvalue. Their signs distinguish restoring and growing perturbations, as derived in Sec.~\ref{sec:normal}. This numerical example differs from the regular initial pattern used in Theorem~\ref{thm:main}.}
\label{fig:flagship}
\end{figure}

The theorem leaves open what happens when some initial projections are negative. A local minimum outside the factual region can prevent a trajectory from reaching the global minimum without contradicting the theorem. Figure~\ref{fig:flagship} illustrates precisely this outcome. Panels (a) and (b) compare identical individuals with identical initial opinions and preferences. Disconnected individuals all correct. On a ring, imitation instead leaves two of the twelve opinions with negative projections. The network permits substantial correction and still preserves a minority of errors.

\section{Why a ring can select a stable state with misinformation}\label{sec:cycle}
\subsection{A constructive comparison with isolated individuals}
A ring provides a simple setting in which local agreement can coexist with a full spread of semantic directions. Let $C_N$ denote the cycle graph on $N\geq5$ nodes, with unit weights on adjacent pairs. Use planar opinions, $n=2$, and choose preferred angles
\begin{equation}
 \psi_i=\frac{\pi}{2N}+\frac{2\pi i}{N},\qquad i=0,\ldots,N-1. \label{eq:anchors}
\end{equation}
Here the node indices are relabeled from zero to describe the cycle. Following the edges takes the opinion directions once around the circle. This is the usual twist or winding configuration of a coupled phase system. Consecutive opinions differ by $\delta=2\pi/N\in(0,\pi/2)$, so neighbors are relatively close even though some opinions oppose the factual direction. The small common offset avoids the exactly parallel and antiparallel factual directions.

\begin{theorem}[Selection of persistent misinformation]\label{thm:main}
For every $N\geq5$ and $h>\kappa>0$, there is an explicit finite threshold $J_{\rm cert}>0$ such that, for $J>J_{\rm cert}$, the initial condition $\theta_i(0)=\psi_i$ converges to a locally exponentially stable equilibrium containing both positive and negative factual projections. This equilibrium is unique in a specified angular neighborhood of the initial pattern. With $J=0$, the same initial condition and preferred directions converge to positive projections at every node. At fixed admissible parameters, both conclusions persist for an open neighborhood of preferred directions, with the initial opinions again set equal to those directions.
\end{theorem}

The threshold is sufficient and need not locate the onset of persistence. Its purpose is to prove a range of finite interaction strengths for which social influence changes the outcome from complete correction to stable coexistence of factual and incorrect opinions. The attractor has higher energy than the factual minimum established by Theorem~\ref{thm:ground}. Thus the collective error is a consequence of dynamical selection, even though a better global compromise exists.

\subsection{The mechanism and the role of the network spectrum}
Imitation resists changes in the angular differences between neighbors. Near the regular pattern, this restoring effect is governed by the cycle Laplacian $L=D-A\in\R^{N\times N}$, where $D=\diag(d_1,\ldots,d_N)$ and $d_i=\sum_jA_{ij}\geq0$ is the weighted degree. The operation $\diag$ constructs a diagonal matrix. The first nonzero eigenvalue of the unit cycle Laplacian is $\lambda_2=2-2\cos\delta>0$. It measures the weakest restoring effect against relative distortions. A common rotation leaves neighbor differences unchanged, but the fixed preferences resist that rotation. These two effects give positive energy curvature near the initial pattern.

Positive curvature alone would establish a local attractor. The proof must also show that the specified initial condition reaches it. We enclose the initial pattern in a small angular neighborhood and show that every boundary point has higher energy than the initial state. Equation~\eqref{eq:dissipation} then prevents the trajectory from crossing the boundary. Strict convexity inside the neighborhood forces convergence to its unique equilibrium. The explicit construction of this energy barrier, together with the coupling threshold, is given in Appendix~\ref{app:certificate}.

\paragraph{Consequence for random initial opinions.}\label{cor:random}
The open neighborhood in Theorem~\ref{thm:main} has positive probability under the independent planar distribution in Eq.~\eqref{eq:law} whenever $0<p<1$. Persistent misinformation therefore does not require exactly tuned preferred directions. This is an existence result at each fixed finite $N$. The probability can depend strongly on the network size, the initial distribution, and the interaction strength; the theorem does not assign its value.

For fixed $N$, $h$, and $\kappa>0$, the selected equilibrium approaches the regular pattern as $J$ increases,
\begin{equation}
 \theta_i^*=\psi_i-\frac{h\sin\psi_i}{J\cos\delta\,\lambda_2}
                 +O(J^{-2}). \label{eq:expansion}
\end{equation}
The superscript $*$ denotes an equilibrium. The remainder $O(J^{-2})$ is bounded in magnitude by a constant times $J^{-2}$ as $J\to\infty$, with the other parameters fixed. The factual field can deform the pattern, but the deformation shrinks as imitation grows stronger. Appendix~\ref{app:expansion} derives this expression and the associated stability estimate.

Fixed preferences make the common phase stable, but they are not essential for persistent error. When $\kappa=0$, an energy barrier around all common rotations of the pattern still keeps at least one opinion negative at every time for sufficiently large $J$. Appendix~\ref{app:tube} states and proves this extension. Its value is to distinguish error sustained by network interactions from error attributed solely to attachment to a fixed preference.

\section{An additional semantic direction changes stability}\label{sec:normal}
The sign of an opinion and its resistance to perturbation answer different questions. The factual projection measures its orientation relative to the reference. Stability measures whether the whole opinion configuration returns after a small displacement. An opinion can be strongly opposed to the factual direction yet belong to an unstable configuration, while a modest negative projection can be part of a stable one.

At an equilibrium, define the signed local field amplitude $\ell_i=\uvec_i^*\cdot\mathbf H_i^*\in\R$. Since the tangent field vanishes there, $\mathbf H_i^*=\ell_i\uvec_i^*$. Let $T_i\in\R^{n\times(n-1)}$ have orthonormal columns spanning the tangent space at $\uvec_i^*$. For small tangent coordinates $\xi_i\in\R^{n-1}$, collect all perturbations in $\xi\in\R^{N(n-1)}$. The constrained energy Hessian, or curvature matrix, is the symmetric matrix $\K\in\R^{N(n-1)\times N(n-1)}$ with blocks
\begin{equation}
 \K_{ii}=\ell_i I_{n-1},\qquad
 \K_{ij}=-JA_{ij}T_i^\top T_j\quad(i\ne j). \label{eq:K}
\end{equation}
The linearized dynamics are $\dot\xi=-\Gamma\K\xi$. We write $\lambda_{\min}(B)$ and $\lambda_{\max}(B)$ for the smallest and largest eigenvalues of any real symmetric matrix $B$. Positive definiteness of $\K$ is equivalent to local exponential stability, and the slowest linear relaxation rate is $\Gamma\lambda_{\min}(\K)$. A negative eigenvalue gives a growing perturbation. A zero eigenvalue requires nonlinear analysis.

Now retain exactly the same planar equilibrium, preferences, and factual vector but embed them in $\R^n$ with $n\geq3$. The new perturbations include rotations out of their common plane. Reordering the tangent coordinates gives
\begin{equation}
 \K_n=\K_\parallel\oplus[\K_\perp\otimes I_{n-2}],\qquad
 \K_\perp=\diag(\ell_1,\ldots,\ell_N)-JA. \label{eq:split}
\end{equation}
Here $\K_\parallel\in\R^{N\times N}$ governs angular changes in the original plane and $\K_\perp\in\R^{N\times N}$ governs one common normal coordinate. The direct sum $\oplus$ denotes independent blocks, and the Kronecker product $\otimes I_{n-2}$ repeats the normal block for each additional coordinate. The decomposition is exact and is derived in Appendix~\ref{app:normal}.

For a cycle equilibrium with $|\theta_i^*-\psi_i|\leq r$, where $r\in(0,\delta/2)$ bounds the angular deviation in real representatives of the cyclic angles, a common normal displacement gives
\begin{equation}
 \lambda_{\min}(\K_\perp)\leq
 h+\kappa-2J[1-\cos(\delta-2r)]. \label{eq:normalbound}
\end{equation}
A negative right hand side proves instability outside the plane. Along the branch selected in Theorem~\ref{thm:main},
\begin{align}
 \lambda_{\min}(\K_\parallel)&=\kappa+O(J^{-1}), \label{eq:parallel}\\
 \frac{\lambda_{\min}(\K_\perp)}{J}&\longrightarrow-2(1-\cos\delta)<0.
 \label{eq:normalasymptotic}
\end{align}
Strong imitation therefore stabilizes the planar pattern against angular distortions but destabilizes its embedding against motion in an extra direction.

This is the geometry shown in Fig.~\ref{fig:flagship}(c). Within a plane, reducing the spread of the ring pattern requires changing neighboring angle differences. Outside the plane, opinions can tilt toward a common new direction and become more mutually aligned. The gain in interaction energy grows with $J$, while the opposing field and preference costs grow with $h+\kappa$. The same geometric distinction is known for unforced spherical consensus \cite{Markdahl2018}; the calculation here evaluates it on the branch selected in the presence of the factual field and fixed preferences.

An exactly coplanar trajectory stays in that plane. Departure requires a nonzero normal perturbation, and the local calculation does not determine its final destination. Randomly sampling preferences throughout a higher dimensional space also changes the fixed fields themselves. That is a different experiment, examined separately in the numerical comparisons.

\section{Numerical evidence for selection and loss of stability}\label{sec:numerics}
The first numerical test follows the construction rather than searching for an equilibrium from an unrelated initial state. We use four cases with $(N,\kappa)=(5,0.75),(8,0.75),(12,0.75),(8,0.25)$, and set $h=\Gamma=1$. In each case, $J=1.2\max(J_{\rm cert},J_{\rm normal})$, where $J_{\rm normal}>0$ is the sufficient threshold for the normal instability bound defined in Appendix~\ref{app:normal}. All four trajectories remain in the region established by the proof and converge to states with both factual signs. Their endpoint residuals are below $8\times10^{-12}$. The residual is the largest remaining tangent field across nodes and measures how nearly stationary a computed state is.

\begin{figure}[!tbp]
\centering
\includegraphics[width=\textwidth]{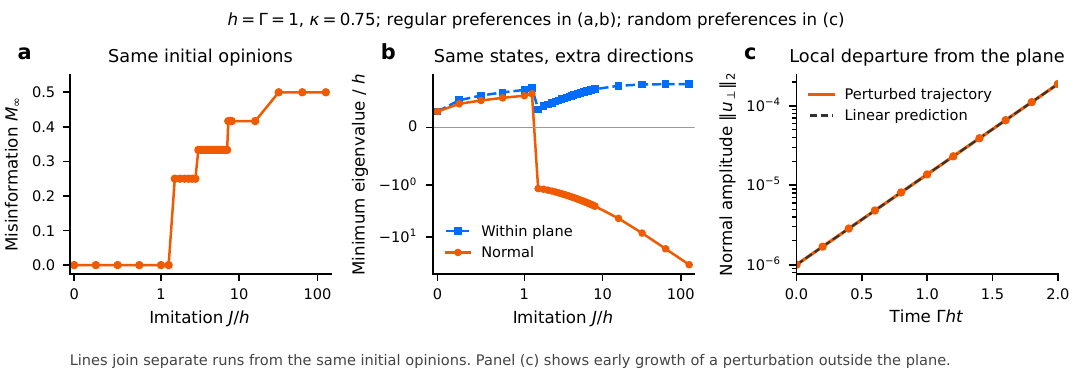}
\caption{\textbf{Stronger imitation can preserve errors and open an instability outside the plane.}
(a) Final misinformation fraction from the regular $N=12$ initial pattern, with $h=\Gamma=1$ and $\kappa=0.75$. Each interaction strength starts from the same preferred directions. The initial rise in $M$ accompanies a large drop in mean factual alignment, whereas later individual steps can reflect crossings of $q_i=0$. (b) Minimum planar and normal Hessian eigenvalues at those endpoints. Positive values indicate local restoring motion in the associated subspace; negative values indicate growth of a perturbation. Coupling axes are linear near zero and logarithmic thereafter, while the eigenvalue axis uses a linear interval around zero and logarithmic scaling at larger magnitude. (c) Growth from a normal perturbation of the random example in Fig.~\ref{fig:flagship}, with initial amplitude $10^{-6}$. Here $\mathbf u_\perp\in\R^N$ collects the opinion components perpendicular to the original plane, and $\|\mathbf u_\perp\|_2$ is its Euclidean amplitude. The early growth agrees with the linear prediction.}
\label{fig:validation}
\end{figure}

Figure~\ref{fig:validation}(a,b) then varies $J$ over 37 values from zero to 128 for the regular $N=12$ pattern with $\kappa=0.75$. Every run starts from the same preferred angles. Panel (a) shows complete correction at $J=0$, an incorrect fraction of $1/4$ at $J=2$, and $1/2$ at $J=32$. The onset between $J=1.25$ and $1.5$ involves a substantial change in the selected configuration. The misinformation fraction rises from zero to $1/4$, while mean factual alignment falls from approximately $0.926$ to $0.440$. The three opinions that become negative previously had projections above $0.9$. They therefore undergo large rotations, rather than small movements across the classification boundary. Since $h>\kappa$, Theorem~\ref{thm:ground} guarantees that the factual equilibrium remains locally stable at both interaction strengths. The change concerns which attractor is reached from the prescribed initial opinions.

Other steps in panel (a) have a simpler origin. In a finite population, $M$ changes in multiples of $1/N$ even when the underlying opinion vectors vary smoothly. From $J=2.75$ to $3$, a single projection changes from approximately $0.002$ to $-0.029$, raising $M$ from $1/4$ to $1/3$. Mean alignment changes only from $0.340$ to $0.327$. This step is consistent with crossing the classification boundary within a family of nearby stable states. Distinguishing these small crossings from a large change in the selected configuration requires inspecting the opinion vectors and the continuous observable $\overline q$, as well as $M$.

Panel (b) asks a different question about the same endpoints. Their planar curvature is positive, whereas the normal curvature becomes negative as interaction grows. A state can therefore retain incorrect opinions as a planar attractor and simultaneously be unstable if an additional direction is allowed. The observed persistence occurs far below the conservative analytical coupling threshold. These repeated initializations locate different selected outcomes at sampled couplings; a continuous bifurcation diagram would require following equilibrium branches between them.

The random example in Fig.~\ref{fig:flagship} provides a second, less regular test. Its preferred directions are sampled uniformly on the circle. Eight of its twelve initial projections are negative. With $J=3$, two remain negative at convergence, approximately $-0.777$ and $-0.617$. Removing the interaction while retaining the preferences gives only positive terminal projections, the smallest approximately $0.662$. The minimum planar and normal curvatures of the coupled equilibrium are approximately $0.426$ and $-2.618$, respectively. Figure~\ref{fig:validation}(c) follows a small perturbation in its most unstable normal mode. The initial growth rate agrees with the predicted rate of approximately $2.618$ for $\Gamma=1$, checking the stability calculation against the dynamics. This example connects the energy argument to an observable outcome of partial correction followed by persistent error.
\FloatBarrier

\section{Random initial opinions and competing outcomes}\label{sec:maps}
The construction proves that a ring can sustain misinformation, but it does not give its typical frequency under random preferences. We next vary preference strength and imitation, follow stable states as parameters change, and compare networks. These experiments address different parts of the same question. The first records what happens from the prescribed initial opinions. The second and third test whether the final state depends on the earlier path or on another current initial condition. The final comparison examines which differences remain when networks share a common spectral scale.

We use the dimensionless parameters
\begin{equation}
 \alpha=\kappa/h\geq0,\qquad \beta=J/h\geq0,
 \qquad \tau=\Gamma h t\geq0. \label{eq:scaled}
\end{equation}
Thus $\alpha$ compares preference with factual influence, $\beta$ compares imitation with factual influence, and $\tau$ measures time in units of the factual response time. All computations below set $h=\Gamma=1$. We use $p=0.5$ and $p=0.1$ in Eq.~\eqref{eq:law}, giving equal probabilities for the two initial hemispheres and a bias toward factual initial opinions, respectively. One independent observation is an entire set of preferred directions on a specified graph. Those directions are reused across parameter values to isolate the effect of changing the forces.

\subsection{Final misinformation and factual alignment}
\begin{figure}[!tbp]
\centering
\includegraphics[width=\textwidth]{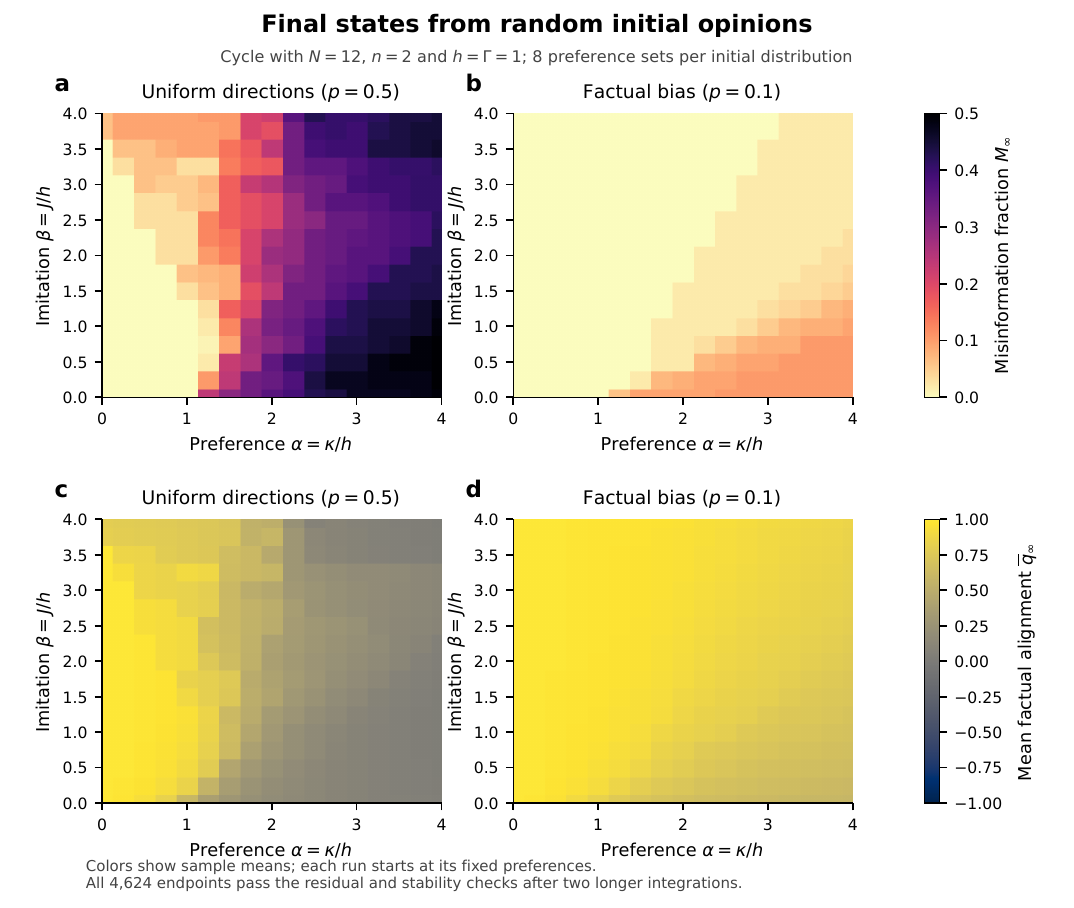}
\caption{\textbf{Terminal states across the parameter plane.}
A planar cycle has $N=12$, $h=\Gamma=1$, and eight independent anchor realizations for each value of $p$. The same anchors are reused on the $17\times17$ grid in $\alpha$ and $\beta$. Panels (a,b) show the sample mean of $M_\infty$; panels (c,d) show the sample mean of $\overline q_\infty=N^{-1}\sum_iq_i$. Colors represent individual parameter cells without interpolation. Anchor realizations are paired across the two initial distributions. The 4,624 endpoints pass the numerical acceptance tests after two separately recorded refinements.}
\label{fig:phase_states}
\end{figure}

Figure~\ref{fig:phase_states} shows the mean terminal misinformation fraction and mean factual projection on the planar $N=12$ cycle. Moving horizontally increases attachment to the preferred directions, while moving vertically increases social imitation. Panels (a,b) give the fraction of opinions that remain negative; panels (c,d) show how closely the population aligns with the factual direction on average. The two columns use different initial distributions. Reading the two rows together distinguishes complete correction from an improvement that leaves some individuals incorrect. The grid has $17\times17$ points, with $\alpha,\beta=0,0.25,\ldots,4$, and eight sets of preferred directions per initial distribution. Each trajectory starts from its own anchors. All 4,624 terminal states meet the numerical residual and spectral tolerances in Appendix~\ref{app:phase_methods}.

The uniform initial distribution retains misinformation in part of the region $\alpha<1$, where the factual signal is strong enough to correct isolated individuals. At $(\alpha,\beta)=(0.75,3)$, two of eight realizations retain negative projections, giving a mean misinformation fraction $5/96\simeq0.0521$ and mean factual alignment $0.8515$. The same anchors correct when $\beta=0$. Positive mean alignment can therefore coexist with persistent error at individual nodes. Across the 68 sampled cells with $\alpha<1$, 22 have a nonzero sample mean of $M_\infty$ for $p=0.5$. The corresponding $p=0.1$ sample has $M_\infty=0$ throughout this part of the grid. The positive probability consequence of Theorem~\ref{thm:main} concerns the existence of such events. The contrast between these two finite samples shows why that existence result must be distinguished from their prevalence under a particular initial distribution.

\FloatBarrier

\subsection{A stable incorrect state can resist correction}
\begin{figure}[!tbp]
\centering
\includegraphics[width=\textwidth]{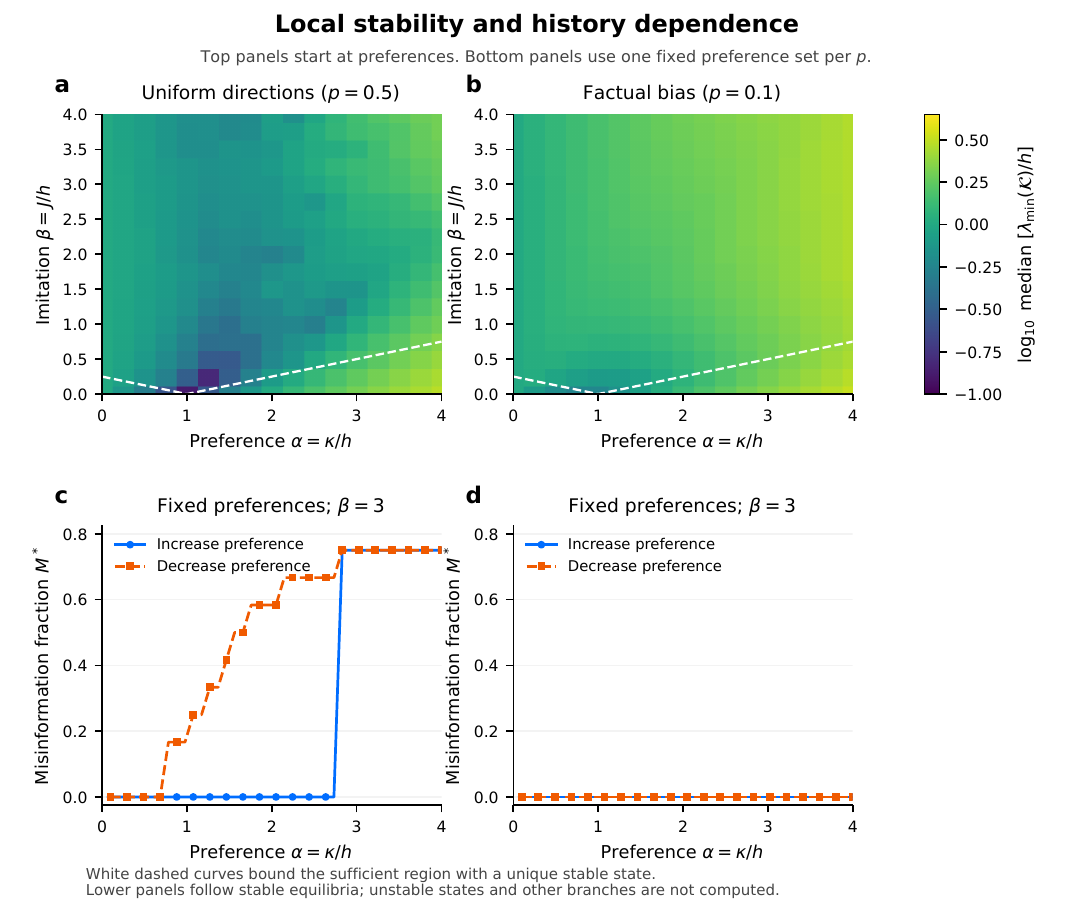}
\caption{\textbf{Spectral margins and attracting continuation branches.}
Panels (a,b) show the logarithm of the median of eight individual minimum Hessian eigenvalues at each cell in Fig.~\ref{fig:phase_states}. The white dashed curves are $\beta=|1-\alpha|/4$; the strict region below them satisfies Proposition~\ref{prop:uniform}. Panels (c,d) follow one fixed anchor realization per $p$ at $\beta=3$. Increasing and decreasing sweeps use 41 values from $\alpha=0.1$ to $4$. All 164 branch points pass the numerical stability test. The two large changes in panel (c) also produce large changes in mean factual alignment, as described in the text. Curves connect sampled attracting states.}
\label{fig:phase_spectra}
\end{figure}

Factual alignment and stability vary separately. Figure~\ref{fig:phase_spectra}(a,b) gives the median of the eight individual minimum tangent Hessian eigenvalues at each cell. Its logarithmic color scale resolves both weak and strong restoring motion. Every displayed state has positive curvature, including cells containing misinformation. In units of $\tau$, the slowest linear relaxation rate of each state is $\lambda_{\min}(\K)/h$. A positive value means that a small perturbation returns toward the selected state, which may itself be incorrect. It does not measure a rate of correction toward the factual direction.

An explicit graph bound identifies regions where competing stable states are excluded. Define $Q=D+A\in\R^{N\times N}$, the signless graph Laplacian, using the weighted degree matrix introduced above. Its largest eigenvalue $\lambda_{\max}(Q)\geq0$ bounds the combined effect of incoming weight and neighbor coupling. The plus sign distinguishes it from the usual Laplacian $L=D-A$.
\begin{proposition}[A spectral condition for a unique stable state]\label{prop:uniform}
If
\begin{equation}
 |h-\kappa|>J\lambda_{\max}(Q), \label{eq:uniform}
\end{equation}
there is exactly one stable equilibrium. It is locally exponentially stable and is the unique global energy minimum. All other equilibria are linearly unstable, and its tangent Hessian satisfies
\begin{equation}
 \lambda_{\min}(\K)\geq |h-\kappa|-J\lambda_{\max}(Q)>0.
 \label{eq:uniformgap}
\end{equation}
If $h>\kappa$, every opinion at the stable equilibrium has positive factual projection. If $\kappa>h$, stable misinformation is possible.
\end{proposition}

Appendix~\ref{app:uniform} proves the sharper condition $\diag(\|\bvec_1\|,\ldots,\|\bvec_N\|)-JQ\succ0$, where $\succ0$ means positive definite, from which Proposition~\ref{prop:uniform} follows. For the unit cycle, $\lambda_{\max}(Q)=4$, so the two certified regions lie strictly below $\beta=|1-\alpha|/4$. These boundaries are shown in Fig.~\ref{fig:phase_spectra}. They are sufficient conditions for a unique stable state, rather than locations where stability is lost. In the region $\alpha<1$, Theorem~\ref{thm:ground} already gives a unique factual global minimum at every $\beta$; the certificate additionally excludes other stable equilibria.

The lower panels in Fig.~\ref{fig:phase_spectra} test history dependence at $\beta=3$. They continue an attracting equilibrium while changing $\alpha$ in small steps, then compare increasing and decreasing sweeps. Unlike the map in Fig.~\ref{fig:phase_states}, a sweep does not restart from the preferred directions at every point. Each sweep first relaxes the opinions from their preferred directions at its own endpoint, then uses the previous equilibrium to initialize the next parameter value. For the fixed $p=0.5$ anchors, the increasing sweep has $M=0$ through $\alpha=2.7325$ and reaches $M=0.75$ at the next sampled value, $\alpha=2.83$. The decreasing sweep retains misinformation down to $\alpha=0.7825$ and reaches $M=0$ at $\alpha=0.685$. The two directions select different misinformation fractions at 21 of the 41 sampled values. The fixed $p=0.1$ example has $M=0$ in both directions. The separation of the traces is numerical hysteresis: the same parameter values and fixed preferences give different final states depending on the direction of the sweep.

The largest changes involve collective reorientation. At the upward step, nine of the twelve opinions change factual sign and $\overline q$ falls from approximately $0.738$ to $-0.253$. At the downward step from $\alpha=0.7825$ to $0.685$, $M$ falls from $1/6$ to zero and $\overline q$ rises from $0.476$ to $0.994$. These changes in a continuous observable show that the two large steps involve more than the discrete classification of opinions close to zero. Increasing $\alpha$ strengthens attachment to the preferred directions relative to the factual field, while imitation couples the resulting rotations. Once an incorrect state has been selected, reducing preference strength need not restore the earlier factual state at the same parameter value. In this example, misinformation persists on the decreasing sweep even at $\alpha<1$, where the isolated individuals would correct.

Both sides of each large step pass the local stability test. The sampled endpoints establish abrupt changes over finite parameter intervals and coexistence over the region where the sweeps differ. Determining whether a followed branch ends, loses stability, or is left after a finite parameter step requires finer continuation. Smaller steps in $M$ can also occur when individual opinions cross the classification boundary while the vectors change only slightly.

\FloatBarrier

\subsection{Several outcomes are possible with the same fixed preferences}
\begin{figure}[!tbp]
\centering
\includegraphics[width=\textwidth]{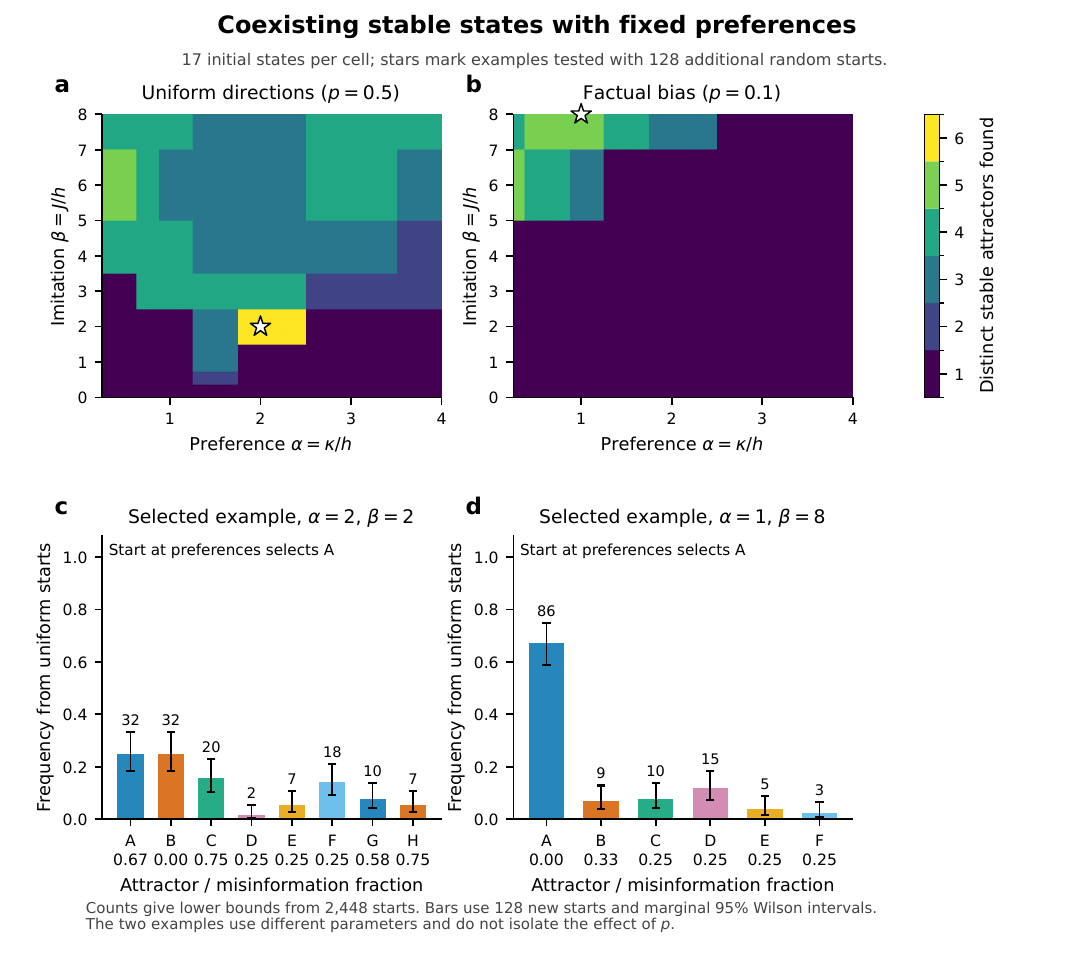}
\caption{\textbf{Several stable endpoints coexist for the same fixed preferences.}
The planar $N=12$ cycle uses one fixed anchor realization per $p$. Panels (a,b) count distinct stable endpoints found from 17 current opinion starts on the parameter grids specified in Appendix~\ref{app:phase_methods}. Stars mark the selected systems used in panels (c,d). At each star, 128 additional independent uniform starts give the displayed selection frequencies; the original 17 starts are excluded from the denominator. Letters identify clusters, their second label gives $M$, and numbers above bars give counts. Error bars are marginal 95\% Wilson intervals. The start at the preferred directions selects cluster A in each panel.}
\label{fig:phase_basins}
\end{figure}

A basin of attraction is the set of current initial states that converge to a particular attractor. To examine these basins in a single fixed system, we hold one set of preferred directions fixed for each $p$ and vary only the current initial opinions. This deliberately relaxes Eq.~\eqref{eq:preparation} for this experiment. Changing an initial opinion here does not change the preference that acts on it during the subsequent dynamics. Figure~\ref{fig:phase_basins}(a,b) reports the number of distinct stable endpoints found from 17 starts at each of 72 parameter pairs. One color therefore counts attractors found in a fixed system, rather than the fraction of incorrect individuals. The 17 search starts consist of opinions equal to the fixed preferences, factual and antifactual consensus, two oppositely oriented twists, and 12 independent uniform configurations. Panels (c,d) then estimate how often two selected systems reach each discovered attractor from 128 additional random starts. All 2,448 search trajectories pass the numerical acceptance tests after three recorded continuations.

The search finds several stable states with different misinformation fractions at the same parameters. At $(p,\alpha,\beta)=(0.5,2,2)$, the original search and 128 additional random starts together identify eight clusters. The start at the preferred directions selects a cluster with $M=2/3$, while another cluster has $M=0$. Each of these two clusters receives 32 of the 128 new random starts. At $(p,\alpha,\beta)=(0.1,1,8)$, six clusters are found; the initial opinions equal to their preferred directions select the $M=0$ cluster, which receives 86 of 128 new starts. Other initial opinions select equilibria with positive misinformation fractions even for these same anchors biased toward the factual hemisphere.

The bars in Fig.~\ref{fig:phase_basins}(c,d) make coexistence visible. Each letter denotes one cluster of nearly identical final vectors. Cluster A is reached from the prescribed preferred directions, but other starts can reach other clusters with different misinformation fractions. Error bars describe uncertainty in the frequency of each cluster, conditional on that fixed system.

The new starts in this experiment are independent uniform current opinion configurations, with the anchors held fixed. Their frequencies describe basins within a selected system, whereas Fig.~\ref{fig:phase_states} varies the anchors and sets $\uvec(0)=\avec$ each time. The cluster count is a lower bound from the finite search. The two displayed systems were selected to illustrate coexistence and have different parameters.

\FloatBarrier

\subsection{What changes with the network, semantic dimension, and population size}
\begin{figure}[!tbp]
\centering
\includegraphics[width=\textwidth]{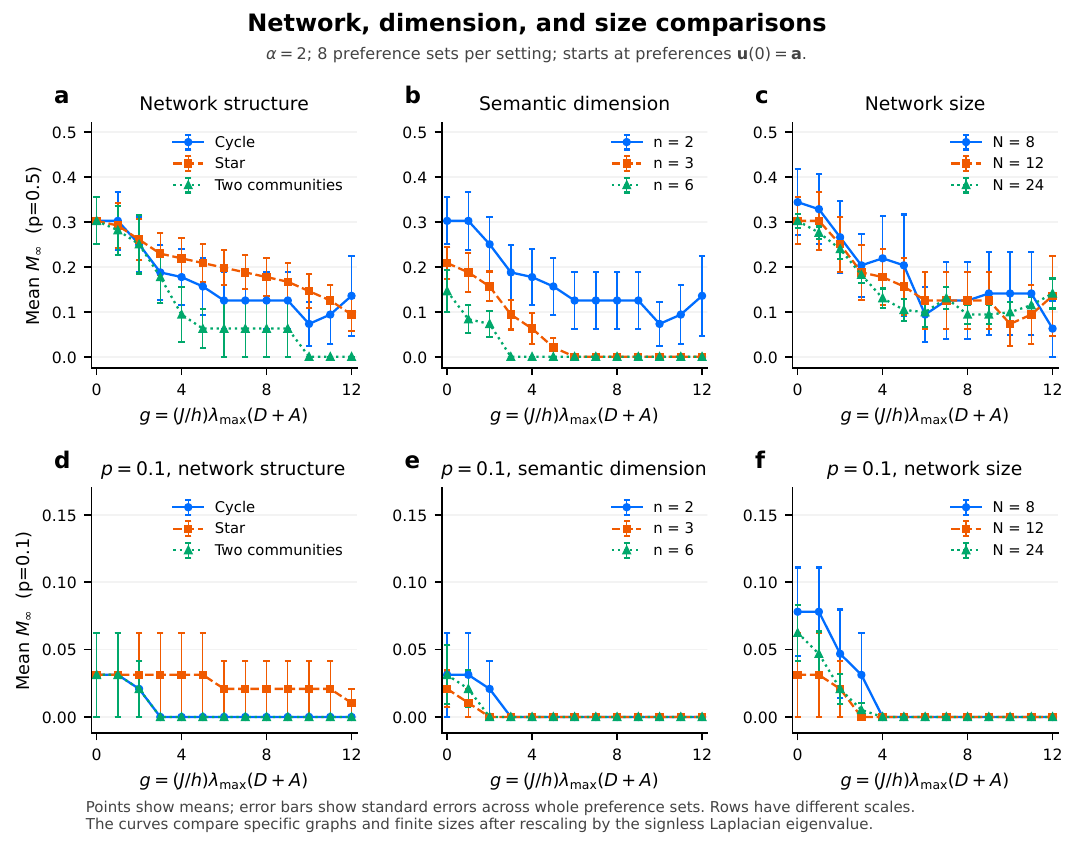}
\caption{\textbf{Graph, dimension, and size comparisons at $\alpha=2$.}
Each point averages eight whole anchor realizations; error bars are the sample standard error of the mean. Panels (a,d) compare three specified graphs at $N=12,n=2$. Panels (b,e) compare $n=2,3,6$ on the $N=12$ cycle. Panels (c,f) compare $N=8,12,24$ cycles at $n=2$. The baseline cycle is shared between comparisons. Anchors are paired between graph structures when $N$ and $n$ agree. Rows use different vertical scales. All 1,456 endpoints pass the numerical tests. Sampling uniform hemispheres at different $n$ changes the distribution of the initial factual projection.}
\label{fig:phase_comparison}
\end{figure}

Figure~\ref{fig:phase_comparison} compares specified graphs, dimensions, and sizes at $\alpha=2$, using eight sets of preferred directions per setting. Here preferences are stronger than the factual field, so some isolated individuals already retain error. These cuts examine variation in correction across model settings; they are distinct from the $\alpha<1$ comparison that isolates error created by interactions. The top row uses $p=0.5$ and the bottom row $p=0.1$. Columns change one model feature at a time. The horizontal coordinate
\begin{equation}
 g=\beta\lambda_{\max}(D+A)\geq0 \label{eq:g}
\end{equation}
uses the same graph scale as the sufficient stability bound. The three graph examples are a cycle, a star, and two complete communities of six nodes joined by one edge, all with $N=12$. Their signless Laplacian radii are $4$, $12$, and approximately $10.4495$, respectively. Equal $g$ therefore compares different raw values of $\beta$.

Panels (a,d) show that terminal misinformation remains distinct after this rescaling. For $p=0.5$ at $g=12$, the sample means are $0.1354$ on the cycle, $0.09375$ on the star, and zero on the graph with two communities. For $p=0.1$, the cycle and the graph with two communities reach zero mean misinformation by $g=3$, while the star retains a nonzero mean at $g=12$. Thus this spectral scale provides a common coordinate but does not collapse the observed graph responses. The same bound can compare the strength of interaction across graphs without predicting which attractor each graph selects.

Panels (b,e) vary semantic dimension on the cycle. The sampled $n=3$ and $n=6$ systems show less misinformation than $n=2$. This comparison uses independently sampled directions in each ambient dimension. Increasing $n$ changes the distribution of the initial factual projection as well as the available directions of motion. The difference is already present at $g=0$, as the exact isolated baseline in Appendix~\ref{app:random_baseline} shows. These simulations therefore complement the normal instability calculation, which instead holds the entire coplanar configuration fixed while adding a perturbation direction.

Panels (c,f) compare cycle sizes $N=8,12,24$. For $p=0.5$ at $g=12$, their sample mean misinformation fractions are $0.0625$, $0.1354$, and $0.1406$, respectively. For $p=0.1$, all three have zero mean by $g=4$. Larger ensembles and sizes are needed to determine how these finite observations scale with population size.

\FloatBarrier

\section{Discussion}
The common formulation makes the relationships between opinion models useful for a correction problem. Its parameter reductions preserve their established mechanisms, while the separate factual reference and specified initial conditions allow agreement, accuracy, and state selection to be compared within the same dynamics.

The results identify a source of persistent misinformation that does not require repulsive social influence, a misleading external signal, or stronger attachment to error than to fact. Under the assumptions of the model, every individual can be correctable in isolation and the collective energy can favor factual orientation globally, while mutual imitation still selects a stable configuration containing error. The ring construction proves this distinction from the same initial opinions and preferences. The numerical maps show related outcomes for randomly sampled preferences at moderate coupling.

The distinction between equilibrium existence and dynamical selection is central. A factual ground state describes what is globally possible under all three influences. An attracting local minimum describes where a trajectory can stop. The energy barrier proves that these answers differ for an explicitly specified initial condition. The additional experiments show two consequences. First, a population can improve its average factual alignment while retaining a minority of incorrect opinions. Second, the same preferences and interaction parameters can support several stable outcomes, selected by earlier states or by the direction in which parameters change.

The abrupt changes in Figs.~\ref{fig:validation}(a) and \ref{fig:phase_spectra}(c) make this selection observable. Their largest steps accompany large changes in continuous factual alignment, while smaller steps can result from individual opinions crossing zero. In the preference sweep, increasing and decreasing attachment lead to different collective outcomes. The incorrect state on the decreasing sweep survives into the region where the factual signal is stronger than preference. Restoring individual correctability can therefore leave a population in an incorrect collective state established earlier. This numerical result connects persistent misinformation to the history of exposure and reinforcement, in addition to the current balance of influences.

Network spectra enter at two different levels. The Laplacian gap controls resistance to relative distortions around the regular ring pattern. The largest signless Laplacian eigenvalue gives a sufficient region with a unique stable state on any finite graph. Outside that region, stability depends on both connectivity and the actual opinion configuration through Eq.~\eqref{eq:K}. The graph comparisons support this distinction. Scaling interaction by a common spectral radius does not remove the observed differences among the ring, star, and graph with two communities. A single structural summary can constrain behavior without determining the selected outcome.

Semantic dimension changes which perturbations exist. For the same coplanar state, an additional direction provides a route that can destabilize a planar local minimum. For independently sampled preferences in different dimensions, both the directions of motion and the distribution of initial factual projections change. The exact isolated baseline separates part of that distributional effect. A theory for generic preferences filling the entire semantic space will need to account for both mechanisms.

\paragraph{Computational methods and data.}
The construction tests and regular coupling cut use Radau integration with analytical Jacobians; the normal perturbation uses DOP853. These are standard implicit and explicit Runge--Kutta solvers, respectively. The parameter maps use fourth order Runge--Kutta integration, with acceptance and refinement details in Appendix~\ref{app:phase_methods}. All perturbations retain the original preferred directions. The ancillary files contain the fixed inputs, terminal states, trajectories, spectra, and reproduction scripts. The new model diagram is conceptual; all other figures use the previously verified numerical data.

\section{Limitations and future directions}
The model isolates competition between factual influence, fixed preferences, and imitation on an undirected network. Its energy arguments rely on reciprocal influence and deterministic dynamics. Directed influence, changing preferences, and noise can alter which states persist. The semantic representation is also an assumption of the theory. A positive factual projection measures orientation toward a reference, while the accuracy of a real statement may involve several independent claims. The finite numerical samples establish coexistence and history dependence in the systems examined. Extending the attracting regions beyond rings and determining the probability of persistent misinformation as population size grows remain theoretical questions.

The main empirical direction is to test the theory in semantic spaces obtained from large language models (LLMs). This requires first establishing a measurable relation between factual accuracy and vector direction. For a specified subject, statements with independently verified factual content and matched false alternatives can be embedded using a chosen model and representation layer. A factual direction can then be estimated from a calibration set and tested on held out statements and paraphrases. This step tests whether the projection distinguishes factual orientation and whether different departures from the reference have a consistent geometric meaning. It also tests the assumption that false statements can be represented on the opposite side of one common boundary.

With that measurement established, controlled LLM agent experiments can compare the same initial opinions and stated preferences with and without network interaction. Agents would receive factual evidence, retain a record of their preferred position, and exchange messages along a specified graph. Their successive statements would be mapped into the calibrated semantic space. The central tests are whether interaction preserves errors that disappear in isolation, whether increasing and decreasing social or preference influence produce different outcomes, and whether allowing additional semantic directions destabilizes a previously persistent pattern. The last test requires holding the factual reference and preferred directions fixed while enlarging the accessible opinion subspace. Effective influence strengths should be estimated from separate response experiments before testing these collective predictions. Such comparisons would show which parts of the directional dynamics describe language based interaction and which require a richer model.

\section{Conclusion}
The model integrates factual influence, persistent preferences, and network imitation in a common directional framework with explicit relations to established opinion models. Its central result is that individually correctable agents can collectively retain misinformation even when the unique global energy minimum is factual. The ring construction proves convergence to an incorrect local minimum from the same initial opinions that correct in isolation. Spectral conditions identify a region without competing stable states, while an additional semantic direction can destabilize the constructed planar pattern. Numerical results show coexisting outcomes, abrupt changes in factual alignment, and history dependent selection. Together, these results explain how attractive social influence can obstruct correction without changing the factual signal or the individuals being compared.

\clearpage
\appendix
\section{Proof of the factual ground state theorem}\label{app:ground}
We prove Theorem~\ref{thm:ground} for fixed local fields $\bvec_i\in\R^n$ with positive factual components $\eta_i=\bvec_i\cdot\fvec>0$. Write the components parallel and perpendicular to the factual direction as
\begin{equation}
 \uvec_i=q_i\fvec+\mathbf y_i,\qquad
 \bvec_i=\eta_i\fvec+\mathbf c_i,
 \qquad \mathbf y_i,\mathbf c_i\in\fvec^\perp.
\end{equation}

\paragraph{Location of the minimum.}
The energy attains a minimum on the compact product of spheres. Reflecting $q_i$ to $|q_i|$ at every node preserves the norm. It strictly decreases the local field contribution if any $q_i<0$. The social interaction contribution cannot increase because $|q_i||q_j|\geq q_iq_j$ and $JA_{ij}\geq0$. Thus every global minimum has $q_i\geq0$ for all $i$. At a node with $q_i=0$, the tangent direction $\fvec$ has energy derivative
\begin{equation}
 -\eta_i-J\sum_j A_{ij}q_j<0.
\end{equation}
Every minimizing projection is therefore strictly positive.

\paragraph{Convexity in the positive hemisphere.}
Parameterize this hemisphere by
\begin{equation}
 q_i=(1-\|\mathbf y_i\|^2)^{1/2},\qquad \|\mathbf y_i\|<1.
\end{equation}
The product of these open balls is convex. The local energy is $-\eta_iq_i-\mathbf c_i\cdot\mathbf y_i$, with second variation along an arbitrary perturbation $\mathbf v_i\in\fvec^\perp$
\begin{equation}
 \eta_i\left(\frac{\|\mathbf v_i\|^2}{q_i}
 +\frac{(\mathbf y_i\cdot\mathbf v_i)^2}{q_i^3}\right)
 \geq\eta_i\|\mathbf v_i\|^2. \label{eq:localconvex}
\end{equation}
For the pair energy $\Phi_{ij}=-q_iq_j-\mathbf y_i\cdot\mathbf y_j\in\R$, let $\mathbf v=(\mathbf v_i,\mathbf v_j)$ denote a joint perturbation. Direct differentiation gives the second directional derivative
\begin{align}
 D^2\Phi_{ij}[\mathbf v,\mathbf v]
 ={}&\left\|\sqrt{\frac{q_j}{q_i}}\mathbf v_i
              -\sqrt{\frac{q_i}{q_j}}\mathbf v_j\right\|^2\nonumber\\
 &+\left[\sqrt{\frac{q_j}{q_i^3}}(\mathbf y_i\cdot\mathbf v_i)
              -\sqrt{\frac{q_i}{q_j^3}}(\mathbf y_j\cdot\mathbf v_j)\right]^2
 \geq0. \label{eq:pairconvex}
\end{align}
Hence the coordinate Hessian of $\E$ is bounded below by $mI_{N(n-1)}$, where the positive constant $m=\min_i\eta_i$ is the smallest factual component of a local field. The minimum already shown to exist in this chart is unique, and there is no other critical point in the chart. At that critical point, the coordinate Hessian and the intrinsic Hessian are related by a nonsingular congruence. The intrinsic Hessian is positive definite, which proves local exponential stability.

\paragraph{Invariance and convergence.}
The factual projections obey
\begin{equation}
 \Gamma^{-1}\dot q_i=\eta_i-q_i(\uvec_i\cdot\bvec_i)
       +J\sum_jA_{ij}\bigl(q_j-q_i\uvec_i\cdot\uvec_j\bigr). \label{eq:qdot}
\end{equation}
At a boundary node with $q_i=0$ and all $q_j\geq0$, the right side is strictly positive. This proves forward invariance of the closed positive region. Let $q_{\min}(t)=\min_iq_i(t)\in[0,1]$ be the smallest factual projection in this region and $B=\max_i\|\bvec_i\|>0$ the largest fixed field magnitude. At an index attaining the minimum, the interaction contribution is nonnegative whenever $q_{\min}\geq0$, since $\uvec_i\cdot\uvec_j\leq1$. Consequently, almost everywhere in time,
\begin{equation}
 \dot q_{\min}\geq\Gamma(m-Bq_{\min}),
 \qquad
 q_{\min}(t)\geq\frac mB+
       \left(q_{\min}(0)-\frac mB\right)e^{-\Gamma Bt}. \label{eq:positivebound}
\end{equation}
The trajectory immediately enters the interior and eventually remains in a compact subset of it. Energy dissipation then places its limit set among the equilibria. The unique critical point in this region is the ground state, so the whole trajectory converges to it. The bound in Eq.~\eqref{eq:localconvex} is a coordinate curvature bound; no particular intrinsic convergence rate is needed here.

\section{An explicit coupling bound for the specified initial opinions}\label{app:certificate}
Use the cycle index $i=0,\ldots,N-1$ and real representatives of the angles near the regular pattern. Write $\theta=\psi+x$, where $\theta,\psi,x\in\R^N$, with the cyclic conventions $x_N=x_0$ and $\psi_N=\psi_0+2\pi$. The energy in these coordinates is
\begin{equation}
 \E(\theta)=-h\sum_i\cos\theta_i
           -\kappa\sum_i\cos(\theta_i-\psi_i)
           -J\sum_i\cos(\theta_{i+1}-\theta_i).
\end{equation}
Let $L=2I_N-A$ be the unit cycle Laplacian, $\delta=2\pi/N$ the regular angle increment, and $\lambda_2=2-2\cos\delta>0$ the smallest nonzero eigenvalue of $L$. Choose a radius $r>0$ for the Euclidean ball of angular displacements satisfying
\begin{equation}
 \delta+\sqrt2 r<\frac\pi2,\qquad
 a_0=\kappa\cos r-\frac{hr}{\sqrt N}>0,\qquad
 r<\frac\pi2-\frac\pi N. \label{eq:rconditions}
\end{equation}
For example,
\begin{equation}
 r=\min\left\{\frac{\pi/2-\delta}{2\sqrt2},
                  \frac{\kappa\sqrt N}{4h},\frac14\right\}
\end{equation}
works for every $N\geq5$ and $h>\kappa>0$. For simultaneous use of the normal bound, include $\delta/4$ in the minimum, giving $r<\delta/2$. The constant $a_0$ bounds the curvature for common rotation. Define the edge cosine bound $c>0$, the field bound $B>0$, and the initial gradient norm $G>0$ by
\begin{gather}
c=\cos(\delta+\sqrt2 r),\quad B=h+\kappa,\quad G=h\sqrt{N/2},\nonumber\\
 J_{\rm cert}=
 \frac{2B^2/a_0+a_0/2+2G^2/(a_0r^2)+h-\kappa\cos r}
 {c\lambda_2}. \label{eq:Jcert}
\end{gather}
All constants are finite and positive. This is a conservative sufficient threshold.

\paragraph{Uniform curvature.}
Let $e_0=N^{-1/2}(1,\ldots,1)^\top\in\R^N$ be the unit vector for common rotation. Decompose any perturbation $z\in\R^N$ as $z=s e_0+y$, with $s\in\R$ and $y\in e_0^\perp$. Here $e_0^\perp$ is the subspace of vectors with zero mean. On $\|x\|_2\leq r$, every edge angle differs from its regular increment by at most $\sqrt2 r$, so the coupling Hessian is bounded below by $JcL$. The $i$th diagonal entry of the field Hessian is $V_i=h\cos(\psi_i+x_i)+\kappa\cos x_i$. Since $\sum_i\cos\psi_i=0$,
\begin{equation}
 e_0^\top\nabla^2\E\,e_0
 \geq\kappa\cos r-\frac hN\sum_i|x_i|\geq a_0.
\end{equation}
The coupling annihilates $e_0$, so the norm of the mixed block is at most $B$. On $e_0^\perp$, the curvature is at least $C=Jc\lambda_2+\kappa\cos r-h$. Young's inequality gives
\begin{equation}
 z^\top\nabla^2\E\,z
 \geq a_0s^2-2B|s|\|y\|+C\|y\|^2
 \geq\frac{a_0}{2}s^2+C_1\|y\|^2,\qquad
 C_1=C-\frac{2B^2}{a_0}. \label{eq:hessianball}
\end{equation}
For $J>J_{\rm cert}$, $C_1>a_0/2+2G^2/(a_0r^2)$, and thus $\nabla^2\E\succeq(a_0/2)I$ on the closed ball, where $I=I_N$ in this angular representation.

\paragraph{Energy barrier and convergence.}
At $x=0$, the anchor and coupling gradients vanish. The vector $h(\sin\psi_i)_i$ is perpendicular to $e_0$ and has norm $G$. Taylor's integral formula, with $x=s e_0+y$ and $\rho=\|y\|\geq0$, gives
\begin{equation}
 \E(\psi+x)-\E(\psi)\geq-G\rho+\frac{a_0}{4}s^2+\frac{C_1}{2}\rho^2.
\end{equation}
On $s^2+\rho^2=r^2$,
\begin{equation}
 \E(\psi+x)-\E(\psi)
 \geq\frac{a_0r^2}{4}-G\rho+
       \left(\frac{C_1}{2}-\frac{a_0}{4}\right)\rho^2
 \geq\frac{a_0r^2}{4}
       -\frac{G^2}{4(C_1/2-a_0/4)}>0. \label{eq:barrier}
\end{equation}
Minimization on the closed ball gives an interior equilibrium. Strong convexity makes it unique there and locally exponentially stable. The trajectory from $\theta(0)=\psi$ cannot reach the boundary because energy is nonincreasing. Its compact sublevel component lies in the interior, where the only critical point is $\theta^*$. The invariance principle for gradient dynamics gives convergence. Strong monotonicity along the segment from $\theta(t)$ to $\theta^*$ also gives
\begin{equation}
 \|\theta(t)-\theta^*\|_2
 \leq e^{-\Gamma a_0t/2}\|\psi-\theta^*\|_2 .
\end{equation}
The trapping region is the connected part of the energy sublevel containing the specified initial state.

\paragraph{Mixed signs and the disconnected control.}
A regular polygon with $N$ vertices has a vertex within $\pi/N$ of the antifactual direction and another within $\pi/N$ of the factual direction. By Eq.~\eqref{eq:rconditions}, their final projections remain strictly negative and positive. Both factual signs therefore persist. At $J=0$, Eq.~\eqref{eq:isolated} applies. The offset $\pi/(2N)$ cannot place an anchor at either $\fvec$ or $-\fvec$. That would require the odd integer $4i+1$ to equal $2N$ times an integer. The prescribed disconnected trajectories all converge to positive projections.

\paragraph{Nearby preferred directions.}
The curvature and margins in the boundary energy are strictly positive on compact sets. They persist uniformly under sufficiently small anchor changes. The perturbed prescribed initial point remains inside the same comparison ball with energy below its boundary. The unique interior minimizer changes continuously, preserving strict negative and positive projections. This gives an open set of preferred directions for which the corresponding initial condition $\uvec(0)=\avec$ selects a state with both factual signs. For $0<p<1$, the independent mixture of uniform hemisphere measures has positive density on the circle, so a product neighborhood in that open set has positive probability. Small admissible symmetric weight perturbations preserve the strict margins at fixed $J$ as well.

\section{Expansion for large interaction strength}\label{app:expansion}
Fix an integer $N\geq5$ and strengths $h,\kappa>0$. Write $\theta_i=\psi_i+s+x_i$, where $s\in\R/(2\pi\mathbb Z)$ is a common phase and $x\in\R^N$ has zero mean, $\sum_i x_i=0$. The small parameter is $\varepsilon=1/J>0$. The energy divided by $J$ is the scalar function $F_\varepsilon$, with social and field parts $W$ and $U$
\begin{align}
 F_\varepsilon(s,x)&=W(x)+\varepsilon U(s,x),\\
 W(x)&=-\sum_i\cos(\delta+x_{i+1}-x_i),\\
 U(s,x)&=-h\sum_i\cos(\psi_i+s+x_i)-\kappa\sum_i\cos(s+x_i).
\end{align}
At $x=0$, the Hessian of $W$ on the subspace of zero mean is $\cos\delta\,L$, which is invertible. The implicit function theorem gives a transverse minimizing branch
\begin{equation}
 x_i(s,\varepsilon)=
 -\frac{\varepsilon h}{\cos\delta\,\lambda_2}
       \sin(\psi_i+s)+O(\varepsilon^2).
\end{equation}
The sine vector is an eigenvector of $L$ with eigenvalue $\lambda_2$, corresponding to one full oscillation around the cycle. Substituting into the energy gives
\begin{equation}
 F_\varepsilon(s,x(s,\varepsilon))
 =W(0)-\varepsilon N\kappa\cos s
 -\frac{\varepsilon^2Nh^2}{4\cos\delta\,\lambda_2}
 +O(\varepsilon^3). \label{eq:reduced}
\end{equation}
The expansion and its fixed finite number of $s$ derivatives are uniform on the compact phase circle. Its term of order two is independent of $s$, since $\sum_i\sin^2(\psi_i+s)=N/2$. Differentiation gives the selected phase $s^*=O(\varepsilon^2)$ and Eq.~\eqref{eq:expansion}.

In the orthonormal decomposition into $e_0$ and its complement, the Hessian of $\E$ has blocks
\begin{equation}
 \K_\parallel=
 \begin{pmatrix}
   \kappa+O(J^{-1}) & O(1)\\
   O(1) & J\cos\delta\,L|_{e_0^\perp}+O(1)
 \end{pmatrix}.
\end{equation}
Rayleigh quotient estimates and bounds from the Schur complement give Eq.~\eqref{eq:parallel}. For large $J$, this branch lies inside the ball used in the proof, so local uniqueness identifies it with the equilibrium selected in Theorem~\ref{thm:main}.

\section{Normal Hessian and geometric energy variation}\label{app:normal}
Consider an equilibrium in a common plane, with $n\geq3$. The angles in this appendix are evaluated at that equilibrium. Choose at each node its unit tangent within the plane and a common orthonormal set of $n-2$ vectors perpendicular to the plane. Then $T_i^\top T_j=\diag(\cos(\theta_i-\theta_j),I_{n-2})$. Substitution into Eq.~\eqref{eq:K} and coordinate reordering prove Eq.~\eqref{eq:split}.

For a unit cycle, use the unit common displacement $N^{-1/2}(1,\ldots,1)^\top$ in the Rayleigh quotient to obtain
\begin{equation}
 \lambda_{\min}(\K_\perp)
 \leq\frac1N\sum_i\ell_i-2J
 =\frac1N\sum_i
 \left[h\cos\theta_i+\kappa\cos(\theta_i-\psi_i)
       +2J\cos(\theta_{i+1}-\theta_i)\right]-2J.
\end{equation}
If $|\theta_i^*-\psi_i|\leq r<\delta/2$, the cyclic edge increments lie in $[\delta-2r,\delta+2r]\subset(0,\pi)$. The lower endpoint is positive, and $\delta+2r<2\delta\leq4\pi/5$ for $N\geq5$. Each cosine is at most $\cos(\delta-2r)$, proving Eq.~\eqref{eq:normalbound}. It is sufficient to impose
\begin{equation}
 J>J_{\rm normal}=\frac{h+\kappa}{2[1-\cos(\delta-2r)]}.
\end{equation}
Taking $J>\max(J_{\rm cert},J_{\rm normal})$ proves planar stability and normal instability simultaneously. Along Eq.~\eqref{eq:expansion}, $\ell_i/J\to2\cos\delta$, and the largest cycle adjacency eigenvalue is $2$, proving Eq.~\eqref{eq:normalasymptotic}.

For an energetic interpretation, let $\mathbf e_\perp\in S^{n-1}$ be perpendicular to the common plane. The angle $\chi\in\R$ parametrizes the configuration path $\uvec_i(\chi)=\cos\chi\,\uvec_i^*+\sin\chi\,\mathbf e_\perp$; it is not dynamical time. Define the scalar field sum $F_{\rm loc}=\sum_i(h\fvec+\kappa\avec_i)\cdot\uvec_i^*$ and edge alignment sum $S_{\rm edge}=\sum_i\uvec_i^*\cdot\uvec_{i+1}^*$. Then
\begin{equation}
 \E(\chi)-\E(0)=(1-\cos\chi)F_{\rm loc}-J\sin^2\chi\,(N-S_{\rm edge}).
\end{equation}
Its second derivative at zero is $F_{\rm loc}-2J(N-S_{\rm edge})$, the Rayleigh numerator for the common normal direction. This direction of decreasing energy is available only when opinions can depart from the original plane. Exact coplanarity remains dynamically invariant.

\section{Persistence without attraction to fixed preferences}\label{app:tube}
Consider the planar unit cycle $C_N$ with integer $N\geq5$, factual strength $h>0$, and no preference attraction, $\kappa=0$. For the initial angles in Eq.~\eqref{eq:anchors}, sufficiently large $J$ keeps at least one factual projection negative at every subsequent time. The invariant region also contains a local energy minimum with negative projections. With $J=0$, the same initial angles converge to the factual direction at every node. We prove these statements by allowing the nearly regular opinion pattern to rotate as a whole.

Let $\mathbf1=(1,\ldots,1)^\top\in\R^N$. Write
\begin{equation}
 \theta=\psi+s\mathbf1+x,\qquad
 s\in\R/(2\pi\mathbb Z),\quad x\in\R^N,\quad \mathbf1^\top x=0,\quad \|x\|\leq r.
 \label{eq:tube}
\end{equation}
The angle $s$ is the common rotation, $x$ is a distortion with zero mean, and $r>0$ bounds its Euclidean norm. With $\delta=2\pi/N$, choose
\begin{equation}
 0<r<\min\left\{\frac{\pi/2-\delta}{2\sqrt2},\frac14\right\}.
 \label{eq:tuberadius}
\end{equation}
The set in Eq.~\eqref{eq:tube} contains all rotations of a nearly regular polygon. It is an embedded tube in the torus $(\R/(2\pi\mathbb Z))^N$ of opinion angles. To see this, suppose two representations differ by $2\pi k$ with $k\in\mathbb Z^N$. Taking differences of coordinates eliminates $s$. Their magnitude is at most $2\sqrt2r<2\pi$, so all entries of $k$ coincide. The zero mean condition then gives the same $x$ and the same phase modulo $2\pi$. The only boundary is $\|x\|=r$.

The social interaction energy divided by $J$ is the scalar function
\begin{equation}
 W(x)=-\sum_i\cos(\delta+x_{i+1}-x_i).
\end{equation}
At $x=0$ its gradient vanishes. Throughout the transverse ball, its Hessian on $\mathbf1^\perp$ is bounded below by
\begin{equation}
 mI,\qquad m=\cos(\delta+\sqrt2r)\lambda_2>0.
\end{equation}
Here $\lambda_2$ is the cycle Laplacian eigenvalue defined above, and $m>0$ is a lower bound for transverse curvature. Thus $W(x)\geq W(0)+m\|x\|^2/2$. The field contribution vanishes at every undistorted phase because $\sum_i\cos(\psi_i+s)=0$, and the cosine Lipschitz bound gives
\begin{equation}
 -h\sum_i\cos(\psi_i+s+x_i)\geq-h\sqrt N\,\|x\|.
\end{equation}
Every boundary point consequently satisfies
\begin{equation}
 \E(\psi+s\mathbf1+x)-JW(0)
 \geq\frac{Jmr^2}{2}-h\sqrt N\,r>0
 \quad\text{if}\quad J>\frac{2h\sqrt N}{mr}. \label{eq:tubebound}
\end{equation}
The initial energy is $JW(0)$. Since energy decreases, its trajectory cannot leave the tube. For every phase $s$, a vertex of the regular polygon lies within $\pi/N$ of the antifactual direction. The distortion at that node is at most $r$, and $\pi/N+r<\pi/2$ by Eq.~\eqref{eq:tuberadius}. At least one projection therefore remains negative at each time.

Minimizing the energy on the compact tube gives an interior minimizer, because its boundary has higher energy than the initial state. The interior is open in the configuration torus, so this is a local minimum of the full energy and contains negative projections. The argument permits degeneracy in its rotational direction. At $J=0$, the initial offset excludes the direction $-\fvec$, and every opinion converges to $\fvec$.

\section{Proof of the sufficient spectral condition}\label{app:uniform}
Set $s_i=\|\bvec_i\|\geq0$, $d_i=\sum_jA_{ij}\geq0$, and $Q=D+A\in\R^{N\times N}$. We prove the stronger statement that positive definiteness of the symmetric matrix
\begin{equation}
 C=\diag(s_i)-JQ\succ0 \label{eq:sharp_certificate}
\end{equation}
implies a unique stable equilibrium and a curvature bound $\lambda_{\min}(\K)\geq\lambda_{\min}(C)$. The notation $C\succ0$ means $z^\top C z>0$ for every nonzero $z\in\R^N$. Let $m_i=s_i-Jd_i$ be a lower bound for the local field magnitude. The diagonal entries of $C$ give $m_i>0$, and the triangle inequality gives $\|\mathbf H_i(\uvec)\|\geq m_i$ at every configuration. Define the normalized local field map $F=(F_1,\ldots,F_N)$ by
\begin{equation}
 F_i(\uvec)=\frac{\mathbf H_i(\uvec)}{\|\mathbf H_i(\uvec)\|}
\end{equation}
The lower bounds $m_i>0$ ensure that this map is defined on the entire product of spheres and takes values in the same space.

For nonzero vectors $x,y\in\R^n$, the identity
\begin{equation}
 \|x-y\|^2=(\|x\|-\|y\|)^2+
 \|x\|\|y\|\left\|\frac{x}{\|x\|}-\frac{y}{\|y\|}\right\|^2
\end{equation}
implies, for any two configurations $\uvec,\mathbf v\in(S^{n-1})^N$,
\begin{equation}
 \|F_i(\uvec)-F_i(\mathbf v)\|
 \leq\frac{J}{m_i}\sum_j A_{ij}\|\uvec_j-\mathbf v_j\|.
\end{equation}
Write $D_m=\diag(m_1,\ldots,m_N)$ and $B=JD_m^{-1}A\in\R^{N\times N}$. Since $D_m-JA=C\succ0$, similarity to the symmetric entrywise nonnegative matrix $JD_m^{-1/2}AD_m^{-1/2}$ gives $\rho(B)<1$, where $\rho(B)$ is its spectral radius, the largest absolute value of an eigenvalue. With $\mathbf1=(1,\ldots,1)^\top\in\R^N$, the weight vector $w=(I_N-B)^{-1}\mathbf1\in(0,\infty)^N$ satisfies $Bw<w$ component by component. Hence $F$ is a strict contraction in the metric
\begin{equation}
 d_w(\uvec,\mathbf v)=\max_i\frac{\|\uvec_i-\mathbf v_i\|}{w_i}.
\end{equation}
The product of spheres is complete in this metric, so $F$ has exactly one fixed point.

At any equilibrium, $\mathbf H_i=\ell_i\uvec_i$ with $\ell_i=\pm\|\mathbf H_i\|\ne0$. If any $\ell_i<0$, a tangent perturbation supported at that node has negative energy curvature and gives a linearly unstable direction. Every stable equilibrium and every global energy minimum must consequently be the unique aligned fixed point. At that point $\ell_i\geq m_i$. For tangent coordinates $\xi_i\in\R^{n-1}$, set $z_i=\|\xi_i\|\geq0$ and $z=(z_1,\ldots,z_N)^\top\in\R^N$. Equation~\eqref{eq:K} gives
\begin{equation}
 \xi^\top\K\xi\geq z^\top(D_m-JA)z
 \geq\lambda_{\min}(C)\|\xi\|^2.
\end{equation}
This proves the refined assertion. Since $s_i\geq|h-\kappa|$, condition~\eqref{eq:uniform} gives
\begin{equation}
 C\succeq\bigl[|h-\kappa|-J\lambda_{\max}(Q)\bigr]I_N,
\end{equation}
which proves Proposition~\ref{prop:uniform}. When $h>\kappa$, the unique minimum is factual by Theorem~\ref{thm:ground}. On the other side, choosing all $\avec_i=-\fvec$ gives the stable antifactual consensus when $\kappa-h>J\lambda_{\max}(Q)$. Thus this second parameter region ensures a unique stable state without requiring factual correctness. The argument does not require graph connectivity.

\section{Exact baseline for random preferred directions without interaction}\label{app:random_baseline}
For $J=0$, Eq.~\eqref{eq:isolated} gives
\begin{equation}
 q_i^*=\frac{1+\alpha r_i}{\sqrt{1+\alpha^2+2\alpha r_i}},
 \qquad r_i=\avec_i\cdot\fvec\in[-1,1].
\end{equation}
The scalar $r_i$ is the factual projection of the preferred direction. Under the continuous sampling law in Eq.~\eqref{eq:law}, the pole and zero field exceptions have probability zero. For $\alpha\leq1$, the terminal projection is positive almost surely. For $\alpha>1$, it is negative precisely when $r_i<-1/\alpha$.

For a uniform direction on $S^{n-1}$, the projection has density
\begin{equation}
 f_n(r)=\frac{\Gamma(n/2)}{\sqrt\pi\,\Gamma((n-1)/2)}
             (1-r^2)^{(n-3)/2},\qquad -1<r<1.
\end{equation}
Here $r\in(-1,1)$ is the projection variable and $\Gamma(\cdot)$ is the gamma function, not the response coefficient in Eq.~\eqref{eq:model}. The hemisphere mixture multiplies this density by $2p$ on $r<0$ and by $2(1-p)$ on $r>0$. Integrating over $r<-1/\alpha$ gives
\begin{equation}
 \mathbb E M_\infty=
 \begin{cases}
 0,&\alpha\leq1,\\[0.3em]
 p\,I_{1-\alpha^{-2}}\!\left(\dfrac{n-1}{2},\dfrac12\right),&\alpha>1,
 \end{cases} \label{eq:random_baseline}
\end{equation}
The expectation is over independently sampled preferred directions. The regularized incomplete beta function is defined, for $x\in[0,1]$ and $a,b>0$, by
\begin{equation}
 I_x(a,b)=\frac{\int_0^x z^{a-1}(1-z)^{b-1}\,dz}
 {\int_0^1 z^{a-1}(1-z)^{b-1}\,dz}.
\end{equation}
For $\alpha>1$, the expression reduces to $2p\arccos(1/\alpha)/\pi$ when $n=2$, and to $p(1-1/\alpha)$ when $n=3$. The expectation is independent of $N$, although the sample fraction fluctuates with the finite anchor realization. It explains why changing semantic dimension also changes the uncoupled baseline in Fig.~\ref{fig:phase_comparison}.

\section{Numerical protocols for the parameter maps}\label{app:phase_methods}
The parameter maps use the classical fourth order Runge--Kutta method, abbreviated RK4. We integrate the angle equations for $n=2$. For $n>2$, we integrate the vector equations in Cartesian coordinates and restore unit length after each complete step. All runs use $h=\Gamma=1$, so the stored time equals the dimensionless time $\tau=\Gamma ht$. With maximum weighted degree $d_{\max}=\max_i\sum_j A_{ij}\geq0$, the default step is
\begin{equation}
 \Delta\tau=\frac{0.25}{1+\alpha+\beta d_{\max}}.
\end{equation}
The terminal residual target is $\max_i\|P_i\mathbf H_i\|\leq10^{-8}$, and the numerical stability threshold is $\lambda_{\min}(\K)>10^{-6}$. Residuals and energy diagnostics are evaluated every 20 steps. Independent checks use smaller time steps, separately written equations integrated with the adaptive explicit solver DOP853, and finite differences along spherical geodesics for higher dimensional Hessians. A geodesic follows a great circle on each unit sphere. These are numerical convergence and consistency checks.

\paragraph{Terminal state maps.}
The $17\times17$ grid uses $\alpha,\beta\in\{0,0.25,\ldots,4\}$, $N=12$, $n=2$, and eight anchor realizations for each $p\in\{0.5,0.1\}$. Two cases that remained unresolved in the original calculation were rerun from their original anchors with half steps, tolerance $10^{-10}$, and maximum time 12,000. Their final times were approximately 8,371.25 and 5,340.91. Independent DOP853 endpoints agree within $9\times10^{-12}$ in maximum node vector distance. These separately stored results replace only the two indexed records when the figures are rendered.

\paragraph{Following stable states as parameters change.}
At $\beta=3$, both directions use the same fixed anchors and 41 equally spaced values of $\alpha$ from $0.1$ to $4$. A sweep starts with the opinions equal to their fixed preferred directions at its endpoint parameter. Each later step starts at the preceding relaxed state with the anchors unchanged. All 164 points pass the residual and spectral criteria.

\paragraph{Basin search and frequencies.}
The search grid is
\begin{align}
 \alpha&\in\{0.25,0.5,0.75,1,1.5,2,3,4\},\nonumber\\
 \beta&\in\{0,0.25,0.5,1,2,3,4,6,8\}.
\end{align}
For each system with fixed preferences and parameters, five designed starts and 12 independent uniform current opinion configurations give 17 trajectories. Along with the residual and spectral tests, the basin calculation requires $\sqrt N\,\mathrm{residual}/\lambda_{\min}(\K)\leq10^{-4}$ as a local convergence diagnostic. Accepted endpoints are linked when their root mean square distance between node vectors, $[N^{-1}\sum_i\|\uvec_i-\mathbf v_i\|^2]^{1/2}$, is below $10^{-3}$. Here $\uvec$ and $\mathbf v$ are two terminal configurations. Connected components of these links define the endpoint groups; the observed diameters within groups are well below that threshold and distances between groups are well above it.

Three unresolved trajectories from the original time 500 search were continued from their saved endpoints with fixed anchors and smaller steps. Their cumulative final times were approximately 704.53, 2,186.52, and 664.43. All 2,448 search endpoints then pass. The displayed examples were selected from the original search by maximizing cluster count, minimizing unresolved count, maximizing the minimum accepted spectral margin, and finally using grid index to break ties. Those examples remained fixed after the refinements.

For each example, 128 fresh uniform current opinion configurations supply the additional starts. These are the only trials in the denominator for the frequency panels. The cluster counts are $(32,32,20,2,7,18,10,7)$ and $(86,9,10,15,5,3)$; each row totals 128. The intervals shown for each cluster are marginal 95\% Wilson intervals.

\paragraph{Graph, dimension, and size comparisons.}
The cuts use $\alpha=2$ and $g=0,1,\ldots,12$. Seven distinct settings supply the three comparisons, with the $N=12,n=2$ cycle reused as their common baseline. Each setting and $p$ uses eight whole anchor realizations. One case required a longer integration; all 1,456 endpoints pass. Error bars use the sample standard deviation across the eight anchor realizations divided by $\sqrt8$. The graph examples are deterministic, so these errors describe anchor variability conditional on each graph.

\paragraph{Changes between sampled states.}
The interpretation of the steps in misinformation fraction uses the saved terminal opinion vectors. We compare the individual factual projections, their mean, and the minimum tangent Hessian eigenvalue at adjacent parameter values. For the increasing and decreasing sweeps, the recorded initial state at each new parameter agrees with the preceding terminal state. These checks use the existing trajectories; the ancillary files include the extracted diagnostics and their script.

\clearpage
\begingroup
\raggedright

\endgroup

\end{document}